\documentclass[nofootinbib,twocolumn,preprintnumbers,amsmath,amssymb,floatfix,pdftex]{revtex4} 
\usepackage[pdftex]{graphicx}
\usepackage{dcolumn}
\usepackage{bm}
\usepackage{multirow}

\begin{document}

\title{Ground-state properties of light $4n$ self-conjugate nuclei in $ab$ $initio$ no-core Monte Carlo shell model calculations with nonlocal $NN$ interactions}

\author{T. Abe$^{1,2}$, P. Maris$^3$, T. Otsuka$^{4,1,5}$, N. Shimizu$^2$, Y. Utsuno$^{5,2}$ and J. P. Vary$^3$}

\affiliation{
$^1$ RIKEN Nishina Center, Wako, Saitama 351-0198, Japan \\
$^2$ Center for Nuclear Study, the University of Tokyo, Hongo, Tokyo 113-0033, Japan \\
$^3$ Department of Physics and Astronomy, Iowa State University, Ames, Iowa 50011, USA \\
$^4$ Department of Physics, the University of Tokyo, Hongo, Tokyo 113-0033, Japan \\
$^5$ Advanced Science Research Center, Japan Atomic Energy Agency, Tokai, Ibaraki 319-1195, Japan
}

\date{\today}

\begin{abstract}

We report $J^\pi = 0^+$ ground-state energies and point-proton radii of $^4$He, $^8$Be, $^{12}$C,  $^{16}$O and $^{20}$Ne nuclei calculated by the {\it ab initio} no-core Monte Carlo shell model with the JISP16 and Daejeon16 nonlocal $NN$ interactions. 
Ground-state energies are obtained in the basis spaces up to seven oscillator shells ($N_{\rm shell} = 7$) with several oscillator energies ($\hbar \omega$) around the optimal oscillator energy for the convergence of ground-state energies. 
These energy eigenvalues are extrapolated to obtain estimates of converged ground state energies in each basis space using energy variances of computed energy eigenvalues.
We further extrapolate these energy-variance-extrapolated energies obtained in the finite basis spaces to infinite basis-space results with an empirical exponential form. This form features a dependence on the basis-space size but is independent of the value of $\hbar\omega$ used for the 
harmonic-oscillator basis functions. 
Point-proton radii for these states of atomic nuclei are also calculated following techniques employed for the energies. 
From these results, it is found that the Daejeon16 $NN$ interaction provides good agreement with experimental data up to approximately $^{16}$O, while the JISP16 $NN$ interaction provides good agreement with experimental data up to approximately $^{12}$C. 
Beyond these nuclei, the interactions produce overbinding accompanied by radii that are too small.
These findings suggest and encourage further revisions of nonlocal $NN$ interactions towards the investigation of nuclear structure in heavier-mass regions. 

\end{abstract}

\maketitle

\section{Introduction 
\label{Sec_1}}

One of the major challenges in low-energy nuclear theory is to understand nuclear structure and reactions from first principles. 
This motivation includes not only the confirmation of existing experimental data, 
but also the prediction of physical observables where experimental information is not yet available.
Moreover, one hopes to extract detailed understanding of a variety of collective nuclear phenomena, described by successful nuclear models, from the underlying microscopic descriptions. For these purposes, a number of {\it ab initio} investigations on nuclear structure have been actively pursued for more than a decade beyond few-body systems with precision few-body calculation techniques 
(see review articles, for example, \cite{Leidemann_2013} and references therein).
This trend is supported by the rapidly growing computational power of supercomputers 
and continuing improvements of {\it ab initio} techniques for nuclear many-body calculations.

Among these {\it ab initio} approaches, we mention that the Green's function Monte Carlo (GFMC) \cite{Carlson:2014vla}, no-core shell model (NCSM) \cite{Barrett_2013}, 
and lattice effective field theory (EFT) \cite{Lee:2008fa} have been applied mainly to $p$- and light $sd$-shell regions. On the other hand, the coupled cluster method \cite{Hagen:2013nca}, in-medium similarity-renormalization-group approach (IM-SRG) \cite{Hergert:2015awm}, self-consistent Green's function (SCGF) theory \cite{Barbieri:2016uib}, and the unitary model operator approach (UMOA) \cite{Shakin_1967,Shakin:1967zza,Suzuki_1994,Miyagi:2019bkl} have been applied mainly to near (sub-)shell-closed medium and heavy nuclei. 
These calculations have provided new insights into nuclear structure that also reflect insights gained from successful traditional shell-model calculations with an assumed inert core in light- and medium-mass region as well as mean-field calculations (including the density functional theory and approaches based on energy-density functionals) covering vast regions of the nuclear chart.   

The NCSM has an established track record for investigating nuclear structure in close comparison with experimental measurements \cite{Navratil_2000_1,Navratil_2000_2}. 
Up to now, there are many NCSM studies of low-lying states around stable $p$-shell nuclei.  
For investigating the exotic nature of loosely-bound (neutron- or proton-rich) nuclei 
and resonant states near decay thresholds showing $\alpha$-clustering structure,   
the effect of coupling to the continuum should be included.  
For such purposes, the no-core Gamow shell model \cite{Papadimitriou:2013ix} and NCSM with the complex scaling method \cite{Papadimitriou:2014bfa} have been investigated mainly in light nuclei. 
For nuclear scattering and reactions, NCSM approaches have been developed such as the NCSM/RGM (no-core shell model/resonating group method) \cite{Quaglioni_2008, Quaglioni:2009mn}, NCSMC (no-core shell model with continuum) \cite{Baroni:2012su, Baroni:2013fe} and the SS-HORSE (single state harmonic oscillator representation of scattering equation) method \cite{Shirokov:2016thl}. 
Furthermore, nuclear densities from the NCSM approach have been used to obtain consistent {\it ab initio} optical potentials for nucleon-nucleus scattering \cite{Burrows:2020qvu,Burrows:2018ggt}. 

For these {\it ab initio} calculations in low-energy nuclear theory 
including the NCSM stated above, all the nucleon degrees of freedom are treated on an equal footing. 
Typically, these calculations tend to be computationally demanding. 
Therefore, alternate ways to reduce the computational cost are needed. 
As one example, we mention the construction of shell-model effective interactions using {\it ab initio} approaches (see, for example, Ref.~\cite{Smirnova:2019yiq} and 
a recent review article \cite{Stroberg:2019mxo}).
In such approaches, the resulting shell-model effective Hamiltonians are diagonalized to obtain physical observables in the conventional shell-model framework with much less computational time and storage or memory. 
Success depends on whether the effects outside the model space can be renormalized properly into the effective interactions between valence nucleons in limited model spaces.

In the NCSM itself, a couple of methods have been proposed to study the structure of heavier nuclei where the standard NCSM cannot be applied even on the current state-of-the-art supercomputers. 
These alternative methods include, for instance, the importance-truncated no-core shell model (IT-NCSM) \cite{IT-NCSM_1, IT-NCSM_2, IT-NCSM_3, IT-NCSM_4, Roth:2013fqa} and symmetry-adapted no-core shell model (SA-NCSM) \cite{SA-NCSM_1, SA_NCSM_6, Dytrych:2016vjy, Launey:2016fvy, Dytrych:2018vkl, McCoy:2020xhp}. 
In particular, the merging of the IT-NCSM and the IM-SRG has been proposed recently \cite{Gebrerufael:2016xih} and has a potential to push further the accessibility of {\it ab initio} approaches. 

The {\it ab initio} no-core Monte Carlo shell model (MCSM) is one of these approaches aiming for successful approximations enabling calculations for heavier nuclei \cite{Abe_2012}. 
So far, the proof-of-principle calculations have been done in relatively small basis spaces for several observables of $p$-shell nuclei in comparison with the results obtained by the no-core full configuration (NCFC) approach \cite{Maris_2009, NCFC_JISP16}, one of already established NCSM methods. For exploratory physics applications, exotic properties in Be isotopes have been studied by the no-core MCSM \cite{Liu_2012}. 
In this work, we extend the no-core MCSM calculations up to $A = 20$ as well as extend  results to several larger basis spaces.  With these extended basis spaces, we perform extrapolations to infinite spaces.  These accomplishments demonstrate the feasibility of the no-core MCSM calculations beyond the $p$ shell on the state-of-the-art supercomputers. 
A major feature of the present work is to investigate to what extent one can describe nuclear structure using only $NN$ interactions without explicit 3$N$ interactions. 
Success in this direction would provide clear advantages for facilitating {\it ab initio} investigations on supercomputers.

The flow of this paper begins with a brief description of the no-core MCSM in Sec.~\ref{Sec_2}. 
We especially focus on how to obtain final {\it ab initio} results from those computed in the finite basis spaces with certain harmonic-oscillator (HO) frequencies. 
In Sec.~\ref{Sec_3}, we introduce the nonlocal $NN$ interactions employed in this paper. 
Then, the numerical results of no-core calculations with these interactions are shown and discussed in comparison with experimental data in Sec.~\ref{Sec_4}. 
Our findings from these results are addressed and some future perspectives are also presented in Sec.~\ref{Sec_4}.
The summary is given in Sec.~\ref{Sec_5}. 
Some numerical results used to obtain final extrapolated results are summarized in Appendices \ref{Appendix_A} and \ref{Appendix_B}.

\section{Method 
\label{Sec_2}}

In this section, we briefly present the formulation of the Monte Carlo shell model (MCSM). 
Originally, the MCSM was developed for the shell-model calculations in the valence space beyond an assumed inert core \cite{Honma_1995, Mizusaki_1996, Honma_1996,Otsuka:1998zz}. 
The MCSM achieved success in describing experimental data (for review articles, see \cite{Otsuka_2001, Shimizu_PTEP_2012, Shimizu_2017}). 
The extension to the no-core MCSM by including major shells for all nucleons appeared straightforward though computationally demanding \cite{Abe_2012, Shimizu_PTEP_2012, Shimizu_2017, Abe:2013hva}. 
Additional key differences emerge when one considers the input  interactions, and the extrapolation to the infinite basis-space limit for the no-core MCSM. 
Here, we describe the general MCSM formulation followed by discussion of the procedures for basis-space extrapolations specific to no-core calculations.

\subsection{Monte Carlo shell model
\label{Subsec_2_1}}

In the MCSM, the intrinsic Hamiltonian is comprised of one- and two-body terms.  
It is written in the second-quantized form as 
\begin{equation}
 {\hat H}_{\rm int} = \sum_{ij} t_{ij} {\hat c}_i^\dagger {\hat c}_j + \frac{1}{4} \sum_{ijkl} {\bar v}_{ijkl} {\hat c}_i^\dagger {\hat c}_j^\dagger {\hat c}_l {\hat c}_k ,
\end{equation}
with nucleon creation and annihilation operators in the spherical HO basis, ${\hat c}^\dagger$ and ${\hat c}$, respectively. The single-particle states are specified by the indices, $i$, $j$, $k$, and $l$. 
The one- and two-body matrix elements are denoted as $t_{ij}$ and ${\bar v}_{ijkl}$. 
Note that the two-body matrix elements are antisymmetrized so that ${\bar v}_{ijkl} = - {\bar v}_{jikl} = - {\bar v}_{ijlk} = {\bar v}_{jilk}$. 
The current version of the MCSM is formulated in the $pn$ formalism (not in the isospin formalism). 
Therefore, the Coulomb interaction can be included naturally within the proton-proton interaction matrix elements. 
Also, spurious center-of-motion effects can, in principle, be alleviated using the Gl{\"o}ckner-Lawson technique as in the conventional shell-model calculations.

With this Hamiltonian, the MCSM state vector is described as a superposition of angular-momentum- and parity-projected basis states, 
\begin{equation}
|\Psi_{JM\pi}^{(N_{\rm b})} \rangle = \sum_{n=1}^{N_{\rm b}} \sum_{K=-J}^{J} f_{nK}^{(N_{\rm b})} {\hat P}_{MK}^{J\pi} |\phi_n \rangle,
\label{Eq.2}
\end{equation}
with the angular-momentum- and parity-projection operator, ${\hat P}_{MK}^{J\pi} = {\hat P}_{MK}^J {\hat P}^\pi$. 
The number of basis states is $N_{\rm b}$, and the coefficient of the basis states is $f$.
In the MCSM, the basis states are expressed as deformed Slater determinants 
and can be written as  
\begin{equation}
|\phi\rangle = \prod_{\alpha=1}^{A} \sum_{i=1}^{N_{\rm sp}} D_{i\alpha} {\hat c}_i^\dagger | - \rangle, 
\label{deformed_Slater_determinant}
\end{equation}
with the particle vacuum $| - \rangle$ (or the assumed inert core in the case of the MCSM with a core), the numbers of nucleons $A$ (or valence particles in the standard MCSM) and of single-particle states in the basis space $N_{\rm sp}$.  
Here, the complex matrix $D$ satisfying the orthonormal condition, $D^\dagger D = 1$,  represents the deformation by mixing HO Slater determinants.  

The matrix elements of $D$ in Eq.~(\ref{deformed_Slater_determinant}) are determined by minimizing energy eigenvalues in stochastic and deterministic ways following the variational principle.
The stochastic sampling of bases is done in a similar way of the auxiliary-field Monte Carlo, 
introducing auxiliary fields by the Hubbard-Stratonovich transformation. 
Candidates for states of the form in Eq.~(\ref{deformed_Slater_determinant}) are generated by different runs of imaginary time evolutions.  
At each step of the basis search, the energy eigenvalues $E$ and coefficients of eigenvectors $f$ are obtained by solving the following generalized eigenvalue problem, 
\begin{equation}
\sum_{nK} \langle \phi_m | {\hat H} {\hat P}^{I\pi}_{MK} | \phi_n \rangle f_{nK} = E \sum_{nK} \langle \phi_m | {\hat P}^{I\pi}_{MK} | \phi_n \rangle. 
\end{equation}
Among these generated candidates, we take the one which gives the lowest energy eigenvalue.
Then, we further minimize the energy eigenvalues by optimizing the complex matrix $D$ in a deterministic way with the conjugate gradient method.
Conventionally, we start with the Hartree-Fock state. 
We increase the number of bases by repeating the basis search in a stochastic and variational way as described above until the energy eigenvalues sufficiently converge. 
The states of the form given by Eq.~(\ref{deformed_Slater_determinant}) that we retain are called our MCSM basis states (``bases'' for short).  We typically retain about 100 of these bases. Therefore, we reduce the diagonalization problem of large and sparse Hamiltonian matrix to a dense Hamiltonian matrix with the ${\cal O}(100)$ bases, enabling the computations in the current state-of-the-art supercomputers. 

\subsection{Extrapolation methods
\label{Subsec_2_2}}

To obtain an estimate of the fully converged energy of a nuclear state within a space with a fixed number of oscillator shells (i.e., fixed $N_{\rm sp}$), 
we also compute the energy variance and  
extrapolate our MCSM results towards vanishing energy variances 
to estimate the exact eigenvalue of the original Hamiltonian for that value of $N_{\rm sp}$ \cite{Shimizu:2010mp}. 
In this extrapolation, the MCSM results are conventionally extrapolated by a second degree polynomial with respect to energy variance. 
We may also use the energy variance as a metric for extrapolating other observables  such as the radius, magnetic-dipole and electric-quadrupole moments, and electromagnetic transition strengths.
Note that the energy-variance extrapolation has also been applied to no-core shell-model (NCSM) calculations \cite{Zhan:2004ct}. 

In our no-core MCSM calculations, we further extrapolate the energy-variance-extrapolated results for sets of $N_{\rm sp}$ towards the infinite basis-space limit 
so as to compare with other {\it ab initio} results and with experimental data. 
In general, there are two parameters defining the HO basis spaces for NCSM calculations.
These are the HO frequency $\omega$ (or equivalently HO energy, $\hbar \omega$) 
of the basis functions and the basis space (or model space) cutoff, $\cal{N}$, which is expressed in terms of the total number of oscillator quanta allowed in the many-body basis states.
In principle, the solution (physical observable in nature) should be independent of the choice of basis space.  For the NCSM we anticipate that an arbitrary observable for a bound state should become independent of the values of $\omega$ as $\cal{N}$ increases towards infinity and the result at this limit can be defined as an {\it ab initio} solution. 
However, due to computational limits, we are restricted to a set of values for $\omega$ and $\cal{N}$. 
Therefore, in order to obtain an {\it ab initio} solution independent of $\omega$ and $\cal{N}$, 
we need to extrapolate our NCSM results. 
One popular extrapolation method is an empirical extrapolation in terms of $\cal{N}$ using fixed value of $\omega$ \cite{Horoi:1998tb, Hagen:2007hi, Forssen:2008qp, Bogner:2007rx, Maris_2009}.
In other approaches \cite{Furnstahl:2012qg, Coon:2012ab, More:2013rma, Furnstahl:2013vda, Furnstahl:2014hca, Konig:2014hma, Wendt:2015nba, Forssen:2017wei} the two parameters $\omega$ and $\cal{N}$ are transformed to (and interpreted as) the infrared (IR) and ultraviolet (UV) cutoffs of the many-body HO basis space, $\lambda$ and $\Lambda$, respectively.
Using these two IR and UV cutoffs, the results are extrapolated to either or both $\lambda \rightarrow 0$ and $\Lambda \rightarrow \infty$. 
By way of comparison, we employ something akin to the former extrapolation method expressed in terms of the number of major shells $N_{\rm shell} = {\cal N} + 1$ with the fixed $\omega$ value, as it is sufficient for our demonstrations of the feasibility of no-core MCSM calculations.  

\section{Calculation details 
\label{Sec_3}}

Here, we present the input interactions and basis-space truncation adopted in the no-core MCSM.
First, we briefly mention the two nonlocal $NN$ interactions employed in this work. 
Next, the parameter setup of the no-core MCSM calculations is described. 
Calculated observables and choice of nuclei are also listed.

\subsection{Interactions
\label{Subsec_3_1}}

In the current {\it ab initio} calculations, one generally employs  $NN$ and 3$N$ interactions based on modern meson-exchange models such as chiral effective field theory ($\chi$EFT) as an input (for review articles, see Refs.~\cite{Epelbaum:2008ga, Machleidt:2011zz}). 
Also, an input interaction is usually softened by renormalization techniques like the similarity renormalization group (SRG) so as to tame the higher momentum components and to facilitate the computation in many-body systems \cite{Hergert:2015awm}.
Extensive efforts are underway to develop consistent interactions in terms of the order-by-order convergence of perturbation and electroweak operators in $\chi$EFT \cite{LENPIC}. 

From the computational demands on both execution time and memory or storage, it is a major challenge to include  3$N$ interactions when progressing to heavier systems.  
It is sufficient for our purposes to defer the inclusion of 3$N$ interactions to a later effort and to focus on the convergence demands already present with the strong $NN$ interaction. Hence, we employ two nonlocal $NN$ interactions, 
the JISP16 \cite{Shirokov_2007} and Daejeon16 \cite{Shirokov_2016} $NN$ interactions. 
Although we treat only $NN$ interactions in these calculations, the results are sufficient to demonstrate the capability of the MCSM technique for no-core calculations and to judge the utility of the adopted nonlocal interactions for a system heavier than those previously examined. 

JISP16 is the abbreviation of the $J$-matrix Inverse Scattering Potential tuned up to the oxygen-16, which is a realistic nonlocal $NN$ interaction constructed through phase-equivalent transformations for the purpose of minimizing the 3$N$-force effects \cite{Shirokov_2007}. 
This interaction is fit not only to two-nucleon scattering data and deuteron properties but also to the properties of light nuclei up to $^{16}$O. 
Summaries of ground-state energies and other properties of light nuclei up through $^{16}$O with the JISP16 interaction are found in Refs.~\cite{NCFC_JISP16, Shirokov_review_JPV382:2014, Shin_2017}

The second interaction we employ is the Daejeon16 $NN$ interaction \cite{Shirokov_2016}. 
This interaction has recently been developed and applied within the NCSM to obtain the low-lying properties of $p$-shell nuclei \cite{Shirokov_2016, Maris:2019etr}. 
In addition, rotational properties of the Be-isotopes have been shown to be emergent properties of NCSM results with Daejeon16 \cite{Caprio:2019lal,Caprio:2019yxh}.  Indeed, emergent rotational properties of Be-isotopes are found to be remarkably similar to collective model descriptions and to results obtained with JISP16 \cite{Caprio:2019mng}.
One of the main differences between the JISP16 and Daejeon16 is the original $NN$ interaction employed as a foundation for the phase-equivalent transformation procedures. 
In the JISP16 $NN$ interaction, a model tridiagonal form of matrix elements in the relative HO representation is used, while for Daejeon16, relative HO  matrix elements of the $NN$ interaction from $\chi$EFT at the N3LO is used \cite{Entem:2003ft}.
For Daejeon16, a SRG transformation is applied to reduce the high-momentum components down to a scale of 1.5 fm$^{-1}$ (the SRG flow parameter).  In the subsequent fits 
to low-lying properties of light nuclei during the construction of phase-equivalent transformations, both JISP16 and Daejeon16 employ the NCFC method to search for optimal parameters of these unitary transformations using selected experimental properties up to $^{16}$O.

\subsection{Basis-space truncation and selection of nuclei
\label{Subsec_3_2}}

In the no-core MCSM, we employ a different truncation scheme for the many-body basis space from that in other NCSM approaches. 
The MCSM truncates the basis space to the single-particle states contained within a number of major shells, $N_{\rm shell}$, while the traditional NCSM truncates the basis space by a cutoff in the sum of excitation quanta from the reference state, $N_{\rm max}$ (for a graphical illustration of the overlaps and differences of the MCSM and NCSM basis spaces, see Fig.~1 of 
Ref. ~\cite{Abe_2012}). 
Thus, configurations taken into account in the calculations with their respective finite basis spaces are different, though, in principle, the results extrapolated into the infinite basis space are expected to be consistent. 

There are two reasons for adopting the $N_{\rm shell}$ truncation for the no-core MCSM.  First, this is a natural truncation for the variational calculations of an initial reference state as adopted in the MCSM either with a core or without a core. 
Second, the nature of $\alpha$ clustering in light nuclei, a focus of planned applications, involves highly collective states 
where many-particle many-hole configurations are expected to dominate. These configurations are favored in the $N_{\rm shell}$ truncation compared with the $N_{\rm max}$ truncation. 

In light of our objectives, we investigate here ground-state properties of $4n$ self-conjugate nuclei, $^4$He, $^8$Be, $^{12}$C, $^{16}$O and $^{20}$Ne.    
Our focus is to confirm the applicability of the no-core MCSM method and establish the foundation for the future investigation of the $\alpha$-cluster structure which appears in $4n$ self-conjugate nuclei and their neighbors.  Perhaps the most celebrated example is the 
 second $0^+$ state of the $^{12}$C nucleus known as the Hoyle state, a highly correlated and collective three-$\alpha$-cluster state near the $\alpha$-decay threshold. 
Therefore, for the current work, we demonstrate the no-core MCSM's capabilities for the energies and point-proton root-mean-square radii of the $J^\pi = 0^+$ ground states of these $4n$ self-conjugate nuclei.

Owing to limited computational resources, the MCSM basis vectors are taken up to 100 states in each basis-space size, except for $N_{\rm shell} = 2$ (where we can obtain fully converged results with respect to the original basis space with less than 100 basis vectors). 
Thus, the obtained results are approximations to the exact ones in the original Hilbert space at each value of $N_{\rm shell}$. 
Therefore, we need to extrapolate our calculated results to obtain converged results for each $N_{\rm shell}$ basis space. 
As mentioned above, we therefore extrapolate energy eigenvalues using a quadratic form with respect to energy variance within each $N_{\rm shell}$ basis space. 
Then, we perform an additional extrapolation of the results from each $N_{\rm shell}$ extrapolation to the infinite value of $N_{\rm shell}$, the full Hilbert space. For this latter extrapolation, we adopt a naive exponential form in $N_{\rm shell}$. There are alternative extrapolation techniques that have been investigated for NCSM calculations, as mentioned above, but these alternative techniques have yet to be adapted for MCSM calculations. We find that the extrapolation procedures we adopt serve our purposes in the present work.

\begin{figure}[htbp]
\includegraphics[width=1.0\columnwidth]{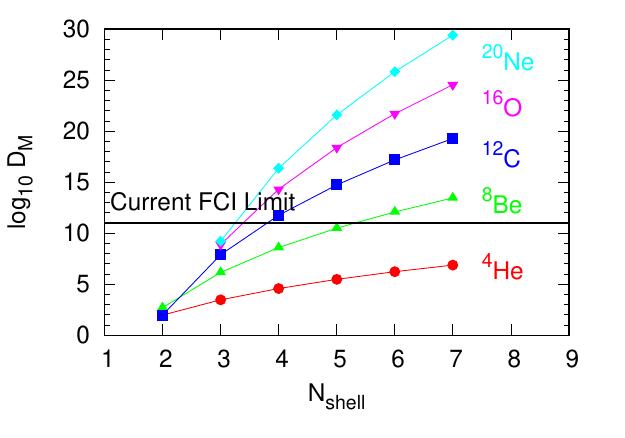}
\caption{$M$-scheme dimensions $D_{\rm M}$ for light nuclei examined in this work as a function of basis space $N_{\rm shell}$. 
The horizontal black line indicates the approximate limit of the dimension 
for full configuration interaction (FCI) calculations using the Lanczos diagonalization method. 
\label{Fig.1}}
\end{figure}

The no-core MCSM basis spaces are taken from $N_{\rm shell} = 2$ to $7$ for $^4$He, $^8$Be and $^{12}$C and $N_{\rm shell} = 3$ to $7$ for $^{16}$O and $^{20}$Ne. 
The many-body basis space dimensions within the $M$-scheme (conserved total angular-momentum projection) for the nuclei examined in this paper are shown in Fig.~\ref{Fig.1}. 
The current limit for full configuration interaction (FCI) approaches, where all many-body basis states within an $N_{\rm shell}$ truncation are retained, is quickly exceeded even by $^8$Be. 
For the HO energy $\hbar \omega$ of basis states, we take the range from  from 10 to 40 MeV in 5 MeV increments so as to cover the optimal values for convergence of our evaluated observables. 

In contrast with the standard NCSM calculations with the $N_{\rm max}$ truncation, 
the spurious center-of-mass (CoM) motion effect generally needs special care using $N_{\rm shell}$ truncation. 
The Gl{\"o}ckner-Lawson technique is employed usually in the no-core MCSM calculations. 
However, the nuclei examined in the present work are deeply bound $4n$ self-conjugate nuclei 
and spurious modes appear sufficiently high in the spectrum, even without the Gl{\"o}ckner-Lawson constraint, that they are well-isolated from the ground state.
Hence, we do not adopt this constraint in the present work. 
As a separate metric of the role of CoM motion in our solutions, we comment that, for $^{20}$Ne in $N_{\rm shell}$ = 7 and $\hbar \omega = 20$ MeV with Daejeon16 interaction, the expectation value of the CoM excitation energy is 424 keV, while that of the intrinsic Hamiltonian is 165.427 MeV. 
This value is comparable to the uncertainty from extrapolations evaluated with the standard least square fit (see Table \ref{table_energy} in Sec.~\ref{Subsec_4_1}). 
Our primary goals in the present work are not affected by this contamination. 
Note that, in the case of excited states, the situation can be quite different from the ground state and we need to handle spurious CoM motion effects carefully, for example, by implementing the Gl{\"o}ckner-Lawson prescription. 

In the shell-model calculations, the Coulomb interaction can be included perturbatively in the isospin formalism or directly by breaking the isosopin symmetry in the $pn$ formalism. 
In this work, we include the Coulomb interaction perturbatively using the isospin formalism in the case of the calculations with the JISP16 $NN$ interaction, and directly using the $pn$ formalism in the Daejeon16 $NN$ case. 
Note that the differences between these two treatments of the Coulomb interaction are negligibly small compared with the uncertainties from extrapolations. 

\section{Results and discussion
\label{Sec_4}}

Here, we present numerical results of the no-core MCSM calculations. 
The calculated energies and radii of light 4$n$ self-conjugate nuclei are shown in comparison of experimental data. 
The differences between the results obtained with the two interactions are discussed.

\subsection{Energy
\label{Subsec_4_1}}

\begin{figure*}[htbp]
\center{
\includegraphics[width=0.90\columnwidth]{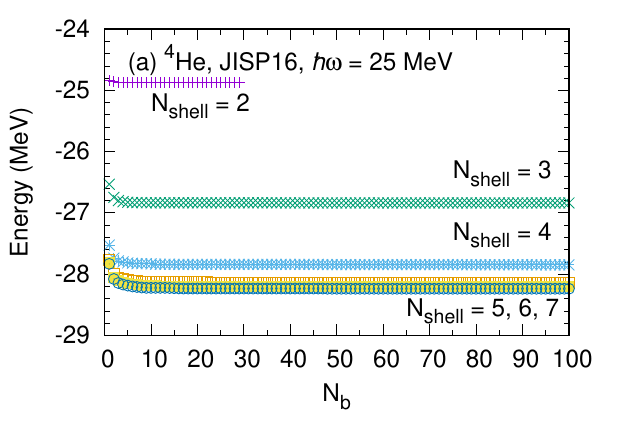}\qquad\includegraphics[width=0.90\columnwidth]{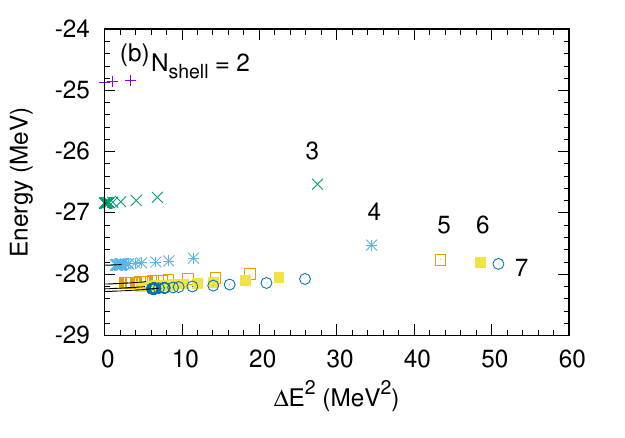}
\includegraphics[width=0.90\columnwidth]{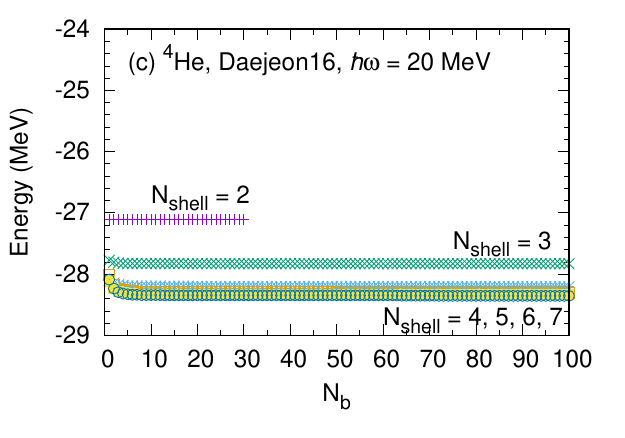}\qquad\includegraphics[width=0.90\columnwidth]{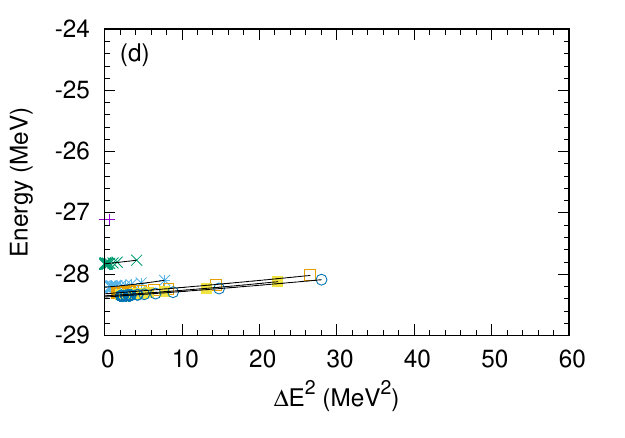}
}
\caption{Convergence of $^4$He ground-state energy with respect to the number of basis states $N_{\rm b}$ in fixed sizes of basis space (left panels) 
and that with respect to energy variances $\Delta E^2$ (right panels).  
The upper panels are the results with the JISP16 $NN$ interaction, 
and the lower panels are for the Daejeon16 $NN$ interaction. 
The Coulomb interaction is included perturbatively in the case of JISP16, 
while directly in the case of Daejeon16.  
The basis space results in $N_{\rm shell} =2$ to $7$ follow a decreasing sequence from the top to the bottom symbols in each panel. 
The results are shown with the optimal value of $\hbar \omega$ (25 MeV for the JISP16 and 20 MeV for the Daejeon16) for the convergence of energy in the largest basis space examined in this work ($N_{\rm shell} =7$).
The black curves in the right panels indicate the fit of a second-order polynomial function to the calculated results. 
See more details in the text. 
} 
\label{Fig.2}
\end{figure*}

Before comparing the calculated results with experimental data, 
we take $^4$He results as an example to be compared with the other {\it ab initio} calculations and experimental measurements.
First, we show the energy-variance extrapolation in the fixed-size of basis space with fixed values of the HO frequency (or HO energy, $\hbar \omega$) 
to approximate the exact eigenenergy in the fixed size of basis space.
Then, we further extrapolate these energy-variance-extrapolated results into the infinite basis space with fixed value of HO energy. 
After seeing the convergence of extrapolated energy by virtue of near-independence of $\hbar \omega$, 
we compare our results with experiment. 
 
In Figure \ref{Fig.2}, we present the ground-state energies of $^4$He with respect to the number of basis states $N_{\rm b}$ in the left panels and energy variance $\Delta E^2$ in the right panels. 
The upper panels are for the JISP16 interaction, while the lower panels are for Daejeon16. 
The other ground-state energies of heavier $4n$ self-conjugate nuclei up to $^{20}$Ne are presented in Appendix \ref{Appendix_A} (Figs.~\ref{Fig.10} and \ref{Fig.11} for JISP16 and Daejeon16 interactions, respectively). 
As the number of basis states increases as depicted in the left panels of Fig.~\ref{Fig.2}, the calculated energy decreases steadily approaching to a converged result for each value of $N_{\rm shell}$.
Note that the MCSM basis vectors are non-orthogonal and the results in $N_{\rm shell} = 2$ are shown with the number of bases up to $N_{\rm b} = 30$ due to the rapid convergence. 
In the $^4$He case, the results shown in the left panels of Fig.~\ref{Fig.2} indicate good convergence with respect to the size of basis space $N_{\rm shell}$.  

In order to improve the estimate of the ground-state energies for each value of $N_{\rm shell}$, we extrapolate our results using energy variance as explained in Sec.~\ref{Subsec_2_2}. 
This process aims to approximate the exact eigenvalue of the full $N_{\rm shell}$ space  from an approximated energy calculated in a truncated basis space defined by $N_{\rm b}$.
This extrapolation step is done using a fit with a second order (in the energy variance) polynomial function (denoted by black solid curves in the right panels of Fig.~\ref{Fig.2}), 
\begin{equation}
  E(\Delta E^2) = a_0 + a_1 \Delta E^2 + a_2 \left( \Delta E^2 \right)^2 + {\cal O} \left( \Delta E^2 \right)^3, 
\label{Eq.5}
\end{equation}
with the energy variance, $\Delta E^2 \equiv \langle \hat{H}^2 \rangle - \langle \hat{H} \rangle^2$. 
The coefficients, $a_0$, $a_1$ and $a_2$, are the parameters for the fit. 
As the number of bases $N_{\rm b}$ increases, the energy variances $\Delta E^2$ approach towards zero indicating the approach to the exact eigenvalue for that value of $N_{\rm shell}$.
In Eq.~(\ref{Eq.5}), the value $a_0 = E(\Delta E^2 \rightarrow 0)$ provides the estimate of eigenenergy of the full basis at that $N_{\rm shell}$.

In the upper panels of Fig.~\ref{Fig.2}, we show the results with the JISP16 $NN$ interaction 
in a comparison with the results with the Daejeon16 $NN$ interaction in the lower panels. 
The comparison of the left panels shows that the convergence with increasing $N_{\rm shell}$ is faster for Daejeon16 than for JISP16.  Looking at the right panels, it is apparent that the Daejeon16 results show typically smaller energy variances than those with JISP16, resulting in the smaller energy difference between the calculated points and the extrapolated value in the case of Daejeon16. 
Thus, the energy obtained with Daejeon16 can be better approximated with a smaller number of bases $N_{\rm b}$ than needed for the energy obtained with JISP16. 
One also notes in the right panels that the Daejeon16 interaction produces smoother trends in the variances compared with JISP16.   
This smoother behavior implies that the Daejeon16 interaction is softer, i.e., closer to perturbative in character, than JISP16. 
These distinctions between results with JISP16 and Daejeon16 are also observed for heavier-mass cases as shown in Figs.~\ref{Fig.10} and \ref{Fig.11} of Appendix \ref{Appendix_A}.  

\begin{figure*}[!ht]
\center{
\includegraphics[width=0.95\columnwidth]{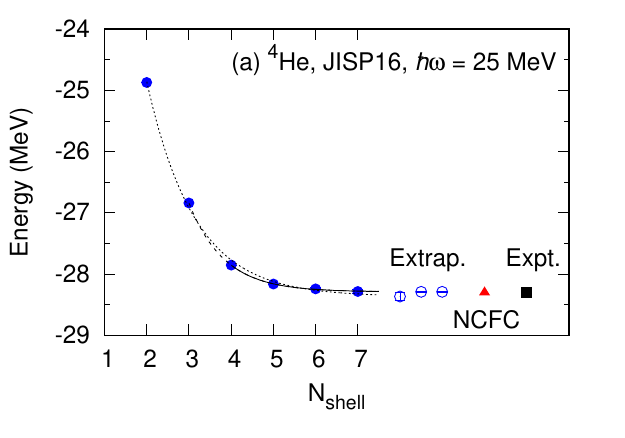}\quad
\includegraphics[width=0.95\columnwidth]{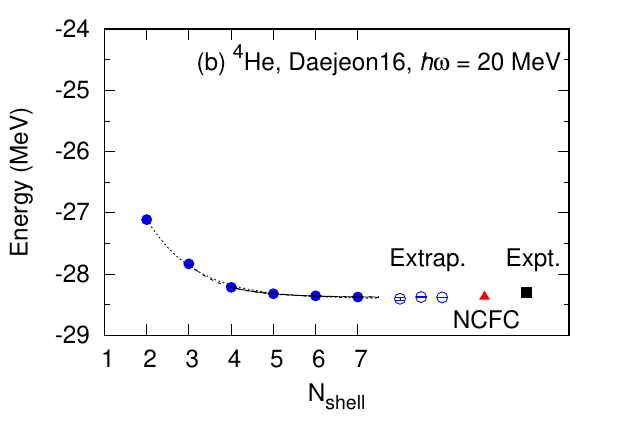}
}
\caption{Convergence of $^4$He ground-state energy with respect to the size of basis space 
with the optimal $\hbar \omega$ value 
obtained by fitting a simple exponential form (\ref{Eq.6}) to results obtained by energy-variance extrapolation within each value of $N_{\rm shell}$. 
The left (right) panel is for the JISP16 (Daejeon16) $NN$ interaction. 
The solid circles indicate the energy-variance-extrapolated energies in each basis-space size. 
The error bars from the energy-variance extrapolation are smaller than the size of the symbols. 
The ``Extrap.'' above three symbols denotes the basis-space-extrapolated energies with the extrapolation errors from the results in $N_{\rm shell} = 2 - 7$, $3 - 7$ and $4 - 7$ from left to right, respectively. 
``NCFC'' denotes the {\it ab initio} results with the JISP16 \cite{Maris_2009, Shirokov_2016} and with the Daejeon16 \cite{Shirokov_2016}.  
``Expt.'' corresponds to the experimental value taken from the compilation of AME2012 \cite{AME2012}.
\label{Fig.3}}
\end{figure*}

Next, we further extrapolate to obtain an estimate in the infinite basis-space limit,  
using the results extrapolated by energy variance in fixed size of basis spaces as shown in Fig.~\ref{Fig.2}. 
This extrapolation towards the infinite basis space is exhibited in Fig.~\ref{Fig.3} (the left panel for the JISP16 $NN$ interaction, and the right for the Daejeon16).  
As discussed above, we adopt a simple extrapolation method in $N_{\rm shell}$ truncation using 
an empirical exponential form, 
\begin{eqnarray}
 E(N_{\rm shell}) = b_0 + b_1 \exp \left( - b_2 N_{\rm shell} \right), 
\label{Eq.6}
\end{eqnarray} 
with three coefficients, $b_0$, $b_1$, and $b_2$, as fit parameters. The value $b_0 \equiv E(N_{\rm shell}  \rightarrow \infty$) provides the estimate in the infinite basis-space limit, when the $E(N_{\rm shell})$ values are taken by energy-variance extrapolated results in each basis-space size with a fixed $\hbar \omega$ value.

In Figure \ref{Fig.3}, we illustrate the extrapolation of ground-state energy of $^4$He at the optimal HO energy with this empirical form (\ref{Eq.6}). 
Here, we take the nearest HO frequency on our chosen grid that provides the lowest energy in the maximum size of basis space ($N_{\rm shell} = 7$ in this work) as the optimal HO energy ($\hbar \omega$ = 25 MeV for JISP16 and 20 MeV for Daejeon16). 
In the figures, there are three curves (solid, dashed, dotted) corresponding to the extrapolations fit to the results in $N_{\rm shell} = 4-7$, $3-7$, and $2-7$, respectively.  
The resultant extrapolated energies in the infinite basis space for these curves are shown with extrapolation uncertainties to the right of the $N_{\rm shell} = 7$ result. For comparison, the exact (experimental) energy is denoted by the black symbol at the rightmost side in the figure. 
As can be seen in the figures, this simple extrapolation works well resulting in rather small uncertainties. In the figures, we also provide the NCSM (NCFC) results using the same interaction, which is in reasonable agreement with ours to within each approach's estimated uncertainty.
Note that the uncertainties from the extrapolations are evaluated simply by using the mean square deviation of the fit coefficients in the least square fitting procedure. 
The basis-space extrapolations for heavier-mass nuclei can be found in Fig.~\ref{Fig.13} of Appendix \ref{Appendix_A}.

\begin{figure*}[!ht]
\center{
\includegraphics[width=0.90\columnwidth]{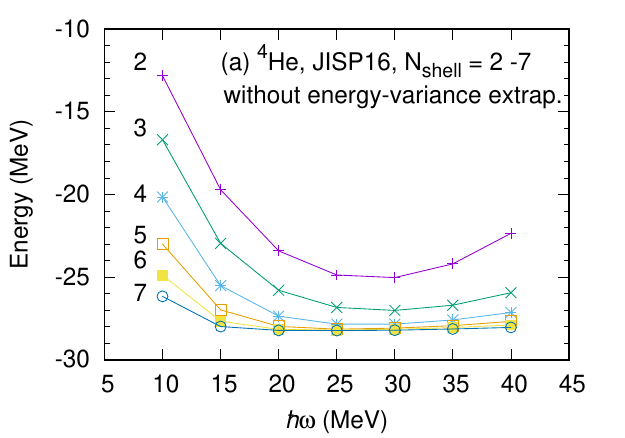}\qquad\includegraphics[width=0.90\columnwidth]{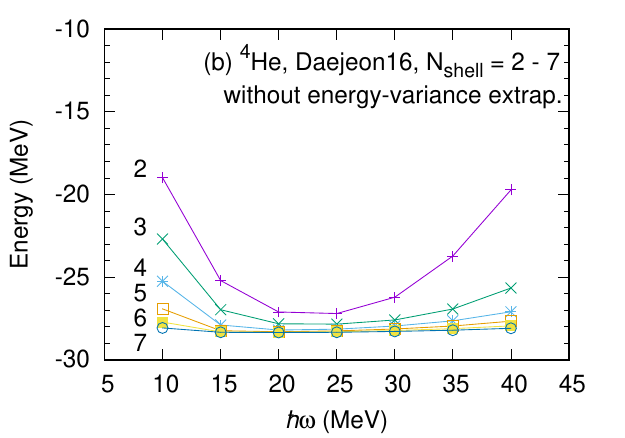}
}
\caption{$^4$He ground-state energy as a function of $\hbar \omega$ 
at the maximum $N_{\rm b}$ in each basis space (i.e. without energy-variance extrapolation).
Results with the JISP16 $NN$ interaction are shown in the left panel 
and those with Daejeon16 $NN$ interaction in the right. 
The numbers in the panels indicate the size of basis space $N_{\rm shell}$. 
The results before the energy-variance extrapolation are presented here. 
  \label{Fig.4}}
\end{figure*}

\begin{figure*}[!ht]
\center{
\includegraphics[width=0.90\columnwidth]{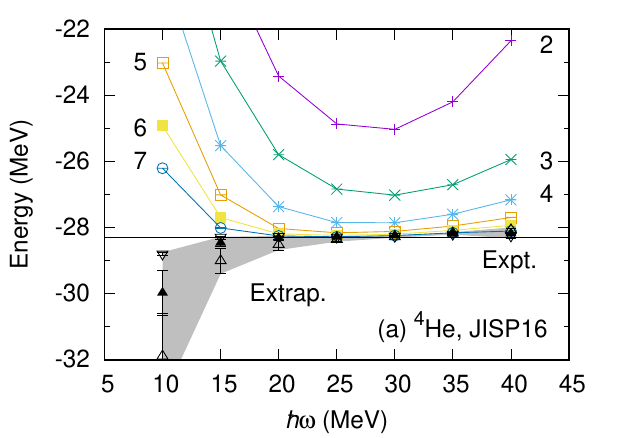}\qquad\includegraphics[width=0.90\columnwidth]{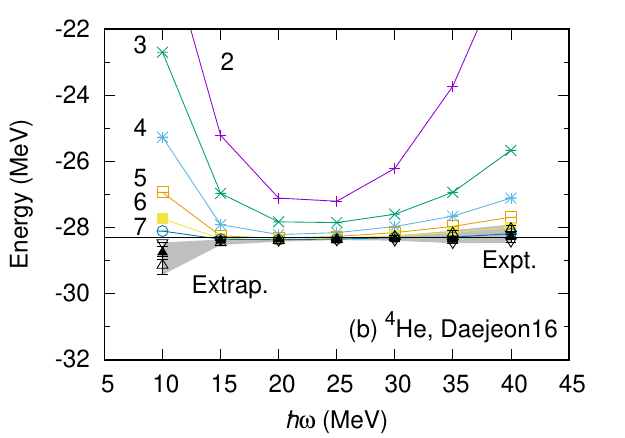}
}
\caption{$^4$He ground-state energy as a function of $\hbar \omega$ including some of the results from Fig.~\ref{Fig.4} on an enlarged scale. 
Results with the JISP16 $NN$ interaction are shown in the left panel 
and those with Daejeon16 $NN$ interaction in the right. 
The horizontal solid line specifies the experimental value. 
The numbers in the panels indicate the size of basis space $N_{\rm shell}$. 
The results after the energy-variance and infinite basis-space extrapolations (``Extrap.'') are  presented here. Note that the errors from the energy-variance extrapolation are smaller than the size of symbols.
The black symbols with gray band are the results extrapolated into the infinite basis-space-size limit. 
The gray bands denote the range of upper and lower edges of three different set of infinite-basis extrapolations. See more details in the text.   
  \label{Fig.5}}
\end{figure*}

In Figure \ref{Fig.4}, we show the results of $^4$He ground-state energy at different HO energies and basis space at the maximum value of $N_{\rm b}$ for each basis space. 
In the figure, the left (right) panel is for the JISP16 (Daejeon16) $NN$ interaction.
From top to bottom, each symbol connected by the solid lines is the result in $N_{\rm shell} = 2 - 7$ prior to energy-variance extrapolation.   
Comparing results between the two interactions, the convergence with respect to the basis-space size $N_{\rm shell}$ is faster in the Daejeon16 case than in the JISP16. 
The optimal $\hbar \omega$ value for the convergence of energy with the Daejeon16 interaction ($\hbar \omega \sim 20$ MeV in $N_{\rm shell} = 7$) is smaller than that with the  JISP16 interaction ($\hbar \omega \sim 25$ MeV in $N_{\rm shell} = 7$). 
These observations reflect the softness of the Daejeon16 interaction compared with JISP16.    
For the corresponding plots of results for heavier-mass cases, we refer to Fig.~\ref{Fig.12} of Appendix \ref{Appendix_A}. 

After our two successive extrapolations in terms of energy variance and basis-space size described above, each at a fixed value of $\hbar \omega$, 
we can now work towards obtaining final results which can be compared with the other {\it ab initio} solutions.
In Fig.~\ref{Fig.5}, we show our final results extrapolated in terms of both energy variance and basis space over a range of $\hbar \omega$ values for $^4$He. 
In the figure, the colored symbols connected by lines are the results extrapolated by energy variances. The uncertainties from the energy variance extrapolations are smaller than the size of symbols if the error bars are not explicitly shown. The black symbols with error bars are the results extrapolated further by the basis-space size to infinity. There are three extrapolated results using the $N_{\rm shell} = 2 - 7$ (upper open triangles), $3 - 7$ (upper solid triangles) and $4 - 7$ (lower open triangles).
The gray band indicates the range of the extrapolations using these three different ranges of basis space size.
For comparison, the horizontal black line is the exact ground-state energy from experiment ($\approx -28.3$ MeV). 
In general, the results extrapolated to the infinite basis-space limit should be independent of the $\hbar \omega$ value, whereas our problem is set in the ($N_{\rm shell}$, $\hbar \omega$) two-parameter space. 
Our extrapolated results seems reasonably flat around the optimal $\hbar \omega$ values. 
The deviations at low and high frequencies indicate the need for larger basis-space sizes to better approximate the infinite basis in those areas. 
In comparison with the JISP16 (left panel) and Daejeon16 (right panel), the Daejeon16 interaction tends to give smaller uncertainties reflecting better convergence with respect to both extrapolations of energy variance for fixed size of basis space and using those extrapolations for extrapolating to the infinite basis-space size. 
As portrayed in Fig.~\ref{Fig.14} of Appendix \ref{Appendix_A}, this relationship of outcomes from the two interactions can also be seen in the heavier-mass cases. 

\begin{figure}[htbp]
\includegraphics[width=1.0\columnwidth]{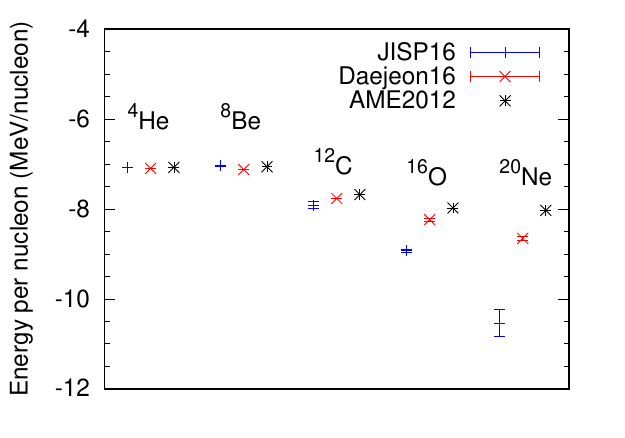}
\caption{Comparison of the extrapolated MCSM results for the energy per nucleon with the JISP16 and Daejeon16 $NN$ interactions to experimental data.
The MCSM results are shown for the extrapolations using the $N_{\rm shell} = 4 - 7$ results with the respective optimal $\hbar \omega$ values. The experimental data are taken from the 2012 Atomic Mass Evaluation (AME2012) \cite{AME2012}. 
\label{Fig.6}}
\end{figure}

Performing the same extrapolations to other heavier $4n$ self-conjugate nuclei from $^8$Be to $^{20}$Ne, we arrive at our final extrapolated results of ground-state energies at their respective optimal HO frequencies and compare these results with experimental data in Fig.~\ref{Fig.6}. 
In Fig.~\ref{Fig.6}, the JISP16 and Daejeon16 results are shown by blue and red symbols, respectively, with estimated error bars from our fit to $N_{\rm shell} = 4-7$ results at their respective optimal $\hbar \omega$ values. 
From Fig.~\ref{Fig.6}, the JISP16 results (blue symbols) yield overbinding at $^{12}$C and beyond.  An improved picture emerges using Daejeon16  (red symbols) with some overbinding still evident when compared with experiments denoted by black symbols. 
Thus there are encouraging trends with the Daejeon16 results compared with the JISP16 results: both improved convergence properties and improved agreement with experiment.

In Table \ref{table_energy}, we also summarize the MCSM results in addition to the NCFC results \cite{Shirokov_2016}.  
Concerning the MCSM results in Table \ref{table_energy}, only the energies extrapolated with the results from $N_{\rm shell} = 4 - 7$ are shown. Additional examples of extrapolated energies and their associated uncertainties can be found in the figures of Appendix \ref{Appendix_A}.  
As shown in Table \ref{table_energy}, the MCSM results are in reasonable agreement with the NCFC results where those are available. 
In nearly all cases, the error bars are overlapping. 
Note that the $\hbar \omega$ values employed in the NCFC calculation are different from 
the $\hbar \omega$ values used in the MCSM calculations. 
The infinite-basis-space extrapolated results are, however, expected to be rather insensitive to $\hbar \omega$ values. 
Our $^{20}$Ne results in Table \ref{table_energy}, 
which are the first-reported {\it ab initio} results for this nucleus with these interactions, 
continue the respective trends in the overbinding that were apparent for these interactions at the upper end of $p$ shell.  
For $^{20}$Ne, we also include the previously unpublished NCFC result with Daejeon16 for comparison. 

\begin{table*}[bt]
\caption{Computed and extrapolated ground-state energies with JISP16 and Daejeon16 $NN$ interactions in comparison with the other {\it ab initio} NCFC results \cite{NCFC_JISP16, Shirokov_review_JPV382:2014, Shirokov_2016, Maris:2019etr, Pieter_in_prep} and experimental data \cite{AME2012}. Note that the NCFC results are obtained with different $\hbar \omega$ values from the MCSM results. 
The procedures for estimating the uncertainties of the MCSM and the NCFC results are different; see Ref.\cite{Pieter_in_prep} for details regarding the NCFC uncertainties. 
  \label{table_energy}}
      \begin{tabular}{ccccccc}
   \hline
   \hline
& & \multicolumn{5}{c}{E (MeV)} \\
   \cline{3-7}
& & \multicolumn{4}{c}{JISP16} & \\
\cline{3-6}
\multirow{2}{*}{Nuclide} & \multirow{2}{*}{$\hbar \omega$ (MeV)} & \multicolumn{3}{c}{MCSM} & \multirow{2}{*}{NCFC} & \multirow{2}{*}{Expt.} \\
& & ($N_{\rm shell} = 7$, $\Delta$E$^2 \ne 0$) &  ($N_{\rm shell} = 7$, $\Delta$E$^2 \rightarrow 0$) & ($N_{\rm shell} \rightarrow \infty$, $\Delta$E$^2 \rightarrow 0$) & & \\
   \hline
$^4$He & 25 & -28.24 & -28.28(3) & -28.29(1) & -28.299(0) & -28.296  \\
$^8$Be & 25 & -53.66 & -54.82(2) & -56.29(4) & -55.5(3) & -56.500  \\
$^{12}$C & 25 & -89.69 & -93.2(3) & -94(1) & -94.8(3) & -92.162  \\
$^{16}$O & 30 & -132.1 & -141.3(7) & -142.9(6) & -145(8) & -127.62  \\
$^{20}$Ne & 30 & -153.4 & -203(7) & -211(6)  & & -160.65  \\
& & \multicolumn{4}{c}{Daejeon16} & \\
\cline{3-6}
$^4$He & 20 & -28.36 & -28.3732(1) & -28.379(6) & -28.372(0) & -28.296  \\
$^8$Be & 15 & -56.45 & -56.774(5) & -56.96(5) & -56.85(7) & -56.500 \\
$^{12}$C &20 & -91.91 & -92.85(4) & -93.19(2) & -92.88(8) & -92.162 \\
$^{16}$O & 20 & -129.4 & -131.25(5) & -131.8(6) & -131.4(7) & -127.62 \\
$^{20}$Ne & 20 & -165.4 & -172.7(2) & -173.0(9) & -168(8) & -160.65  \\
   \hline
   \hline
  \end{tabular}
\end{table*}

\subsection{Radius
\label{Subsec_4_2}}

Following similar procedures, we have also evaluated the point-proton root-mean-square (rms) radii of these light $4n$ self-conjugate nuclei. 
However, as distinct from the evaluation of the ground-state energy, we have not fully applied the extrapolation with energy variances at fixed values of $N_{\rm shell}$.
Although the radius seems to be increasing as the energy variance is approaching to zero as shown in the figures of Appendix \ref{Appendix_B}, it is difficult to perform robust extrapolations for all the radii with respect to energy variance due to the nonmonotonic behavior of the calculated radii.  
Contrary to the case of the energy, even though we can fit some of the results in relatively small basis-space sizes with a linear function such as those done in Ref.~\cite{Abe_2012},  
we refrain from fitting the cases with manifestly noisy data.  Instead, for those cases we merely attempt to capture the general tendency of calculated results with these two nonlocal interactions.
Here what we would like to address is the capability of the MCSM calculations themselves and simply show the current status of the computations on the state-of-the-art supercomputers. 

\begin{figure*}[t]
\center{
\includegraphics[width=0.90\columnwidth]{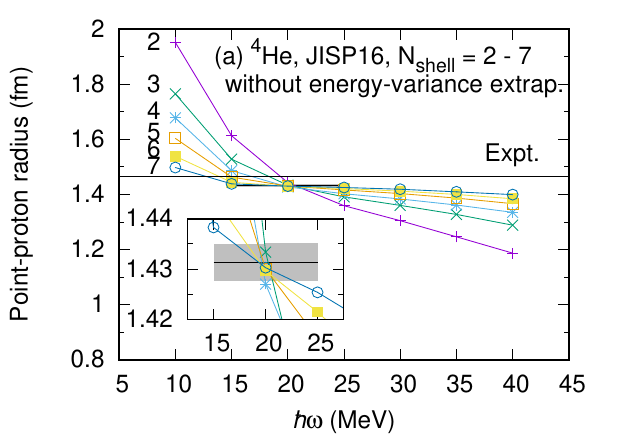}\qquad\includegraphics[width=0.90\columnwidth]{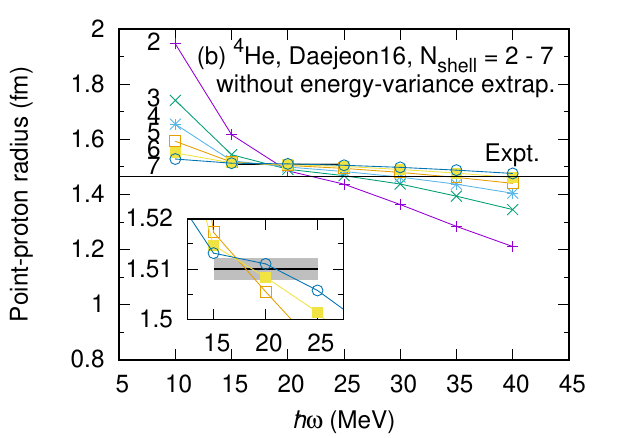}
}
\caption{Point-proton root-mean-square (rms) radius of the $^4$He ground state as a function of $\hbar \omega$. 
Results with the JISP16 $NN$ interaction are shown in the left panel 
and those with Daejeon16 $NN$ in the right panel. 
The MCSM results before the energy-variance extrapolations are plotted here. 
The numbers in the leftmost side of figures specify the size of basis space $N_{\rm shell}$. 
The horizontal black bar with gray band denotes the average value of the rms radius with the uncertainty around the optimal HO frequency range (three successive points with least sensitivity to the HO frequency).
The horizontal black line stretching across each full panel is the experimental value taken from the ADNDT2013 \cite{ADNDT2013}. 
The inset is an enlarged drawing of the average of the MCSM results around the optimal $\hbar \omega$ values. 
  \label{Fig.7}}
\end{figure*}

In Figure \ref{Fig.7}, we show the extrapolation of no-core MCSM results for the $^4$He nucleus into the infinite basis-space size without the energy-variance extrapolation. 
As $N_{\rm b}$ increases, the results reveal a region of significant independence of $\hbar \omega$. 
To estimate the radius in the infinite basis-space limit, 
we simply average the results around the inflexion point of curves connecting the results in the same basis-space size. 
To get the average, we take two or three points in different $\hbar \omega$ values depending on the location of the inflexion point.  
In the figure, the horizontal bar denotes the average of those results around the inflexion point 
and the band portrays the uncertainty.  
The black horizontal line is for the point-proton radius extracted from the experimental charge radius. 
The expectation value of the squared charge radius $\left< {\hat r}^2 \right>_{\rm ch}$ measured experimentally is related 
to that of the squared point-proton radius $\left< {\hat r}^2 \right>_{\rm pp}$ calculated theoretically through
\begin{equation}
  \left< {\hat r}^2 \right>_{\rm ch} \simeq \left< {\hat r}^2 \right>_{\rm pp} + \left< {\hat R}^2_{\rm p} \right> + \frac{N}{Z} \left< {\hat R}^2_{\rm n} \right> + \frac{3}{4m_{\rm p}^2}, 
\label{Eq.7}
\end{equation}
with the squared charge radius of proton $\left< {\hat R}^2_{\rm p} \right>$, the squared charge radius of neutron $\left< {\hat R}^2_{\rm n} \right>$, 
the proton mass $m_{\rm p}$, and the proton (neutron) number $Z$ ($N$) \cite{Friar:1997js}.
The last term in Eq.~(\ref{Eq.7}) is known as the Darwin-Foldy term associated with the relativistic correction in natural units.
Note that the spin-orbit corrections and the higher-order effects such as the meson-exchange currents are not shown,  
because those are expected to be smaller than the resolution of experimental measurements.
The values of $\left< {\hat R}^2_{\rm p} \right>$ and $\left< {\hat R}^2_{\rm n} \right>$ are taken from the ADNDT2013 compilation \cite{ADNDT2013} 
as $\left< {\hat R}^2_{\rm p} \right>^{1/2} = 0.8783(86)$ fm and $\left< {\hat R}^2_{\rm n} \right> = - 0.1149(27)$ fm$^2$, respectively. 
The squared point-proton radius is calculated with
\begin{equation} 
\left< {\hat r}^2 \right>_{\rm pp} 
= \frac{1}{Z} \sum_{i=1}^Z \left| {\hat r}_i - {\hat r}_{\rm CM} \right|^2, 
\label{Eq.8}
\end{equation}
and the CoM coordinate, ${\hat r}_{\rm CM} = \sum_{i=1}^A {\hat r}_i /A$.  
In Eq.~(\ref{Eq.8}), both one- and two-body operators are involved \cite{Caprio:2014iha}. 
We directly evaluated the expectation values of these operators 
similar to the calculations of the ground-state energy.
The agreement with the experimental value for $^4$He is reasonable both for the JISP16 and Daejeon16 $NN$ interactions.
The JISP16 (Daejeon16) result gives a little bit smaller (larger) value compared with the result derived from experiment.
The calculated point-proton radii for nuclei heavier than $^4$He can be found in Appendix \ref{Appendix_B}. 

\begin{figure}[htbp]
\includegraphics[width=1.0\columnwidth]{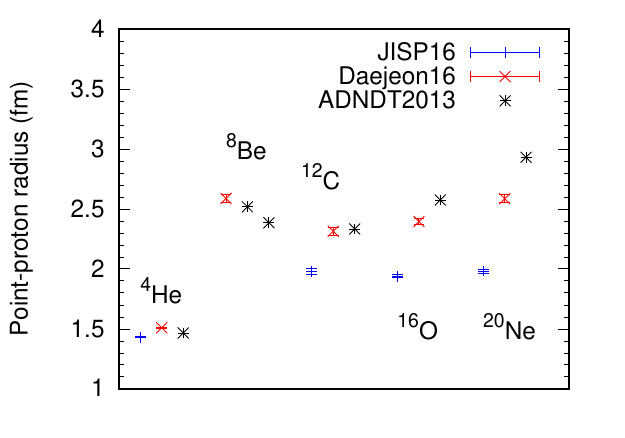}
\caption{Comparison of the MCSM results of point-proton rms radius with the JISP16 and Daejeon16 $NN$ interactions to experimental data.
The MCSM results with error bars are shown for the averages around the optimal $\hbar \omega$ values (three successive points with the least sensitivity to the HO frequency at $N_{\rm shell} = 7$). 
The experimental data are taken from the ADNDT2013 \cite{ADNDT2013}. 
In the case of $^8$Be, the radii for the ground states of the neighboring nuclei, $^{7,9}$Be $\frac{3}{2}^-$ states (the left black symbol for $^7$Be and the right for $^9$Be) are taken as reference radii, because of the missing experimental value of $^8$Be. The result with JISP16 for the $^8$Be radius are not shown in the figure (see more details in the text and the caption of Table.~\ref{table_radius}). 
\label{Fig.8}}
\end{figure}

Similar to the ground-state energies summarized in Fig.~\ref{Fig.6}, 
we show the calculated point-proton root-mean-square radii of ${}^4$He, ${}^8$Be, ${}^{12}$C, ${}^{16}$O and ${}^{20}$Ne nuclei with the JISP16 and Daejeon16 $NN$ interactions in Fig.~\ref{Fig.8}. 
The results shown in the figure are the averages around optimal $\hbar \omega$ values.  
As seen in the figure, the Daejeon16 results appear closer to the experimental values than the JISP16 results.
This behavior is correlated with the corresponding results for energies in Fig.~\ref{Fig.6}: 
that is, more binding correlates with smaller radii.
However, even using the Daejeon16 interaction, theoretical results give smaller radii than those derived from experiment.
Similar tendencies are also found in the Hartree-Fock many-body perturbation theory 
(HF-MBPT) calculations for $^4$He and $^{16}$O with the JISP16 interaction \cite{Hu:2016txm}. 
In Table \ref{table_radius}, the calculated point-proton rms radii are summarized in comparison with those extracted from experimental charge radii. 
For $^{12}$C, our current results are in good agreement with NCSM calculations of the point-proton rms radius \cite{Shirokov_2016} using the same interactions.

Note that $^8$Be is a special case since it is unbound with respect to the two-$\alpha$-decay threshold in nature.  
As seen in Table \ref{table_energy}, the experimental value of the $^8$Be ground-state energy is $-56.500$ MeV, 
while twice the experimental $^4$He ground-state energy is $-56.592$ MeV. 
The no-core MCSM results with the JISP16 interaction also suggest the unbound nature of $^8$Be with respect to the two-$\alpha$-decay threshold ($-56.29$ MeV for the $^8$Be ground-state energy and $-56.58$ MeV for twice the $^4$He ground-state energy). 
The results with Daejeon16, however, tend to be weakly bound with respect to the two-$\alpha$ decay threshold ($-56.96$ MeV for the $^8$Be ground-state energy and $-56.758$ MeV for twice the $^4$He ground-state energy).
Note that this weakly-bound trend with Daejeon16 is supported by the NCFC calculations as shown in Table \ref{table_energy}.
  
This issue also impacts the calculated point-proton radius for $^8$Be.  
As shown in Fig.~\ref{Fig.18} of Appendix B, the convergence pattern of the $^8$Be point-proton radius differs between the JISP16 and Daejeon16 interactions. 
As the basis-space size increases, we did not observe the inflexion point in the case of JISP16 
at least at the range of $\hbar \omega$ we examined.  
We did not extrapolate the point-proton radius in this case.     
Since the $^8$Be ground state calculated with JISP16 is unbound, 
the radius, in principle, should diverge as the basis space goes to infinity. 
These results tend to support that since the point-proton radius with Daejeon16 
suggests a much better convergence pattern than the $^8$Be point-proton-radius results 
above for JISP16. 

\begin{figure}[htbp]
\includegraphics[width=1.0\columnwidth]{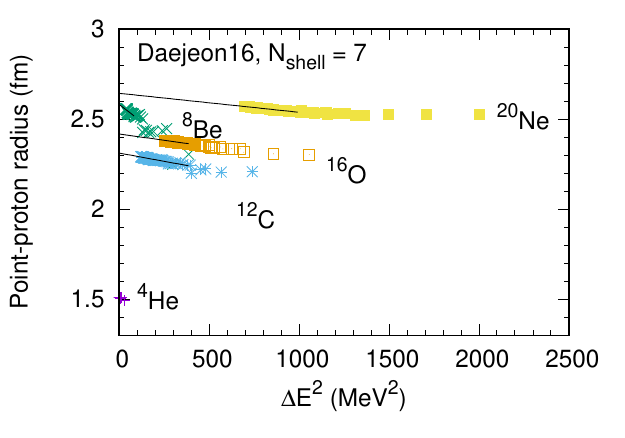}
\caption{Energy-variance extrapolation of point-proton rms radii using the results with the Daejeon16 $NN$ interaction. 
The calculated radii are from $N_{\rm shell} = 7$ results around the optimal HO frequencies ($\hbar \omega =$ 10 MeV for $^4$He and 15 MeV for $^8$Be, $^{12}$C, $^{16}$O and $^{20}$Ne). 
The solid lines are the fits to the computed radii by the linear functions. 
The range of the fit is the region where the points and line overlap. 
\label{Fig.9}}
\end{figure}

As noted above, these calculated radii are, until now, without the energy-variance extrapolation. 
It is still an open question to what extent the calculated radii underestimate the experimental data, quantitatively. 
To attempt a partial answer to this question, we extrapolate the results using energy variance for calculated results that are not excessively noisy.  
The energy-variance extrapolations of the results with the Daejeon16 interaction in $N_{\rm shell} = 7$ around the respective optimal $\hbar \omega$ values are shown in Fig.~\ref{Fig.9}. 
From this figure, the energy-variance extrapolation increases the rms radii slightly relative to their $N_{\rm shell} = 7$ results ($\approx 0.07$ fm increase for $^{20}$Ne at most), 
but does not fill the gap between the results and experimental data.  
Simultaneous reproduction of energy and point-proton rms radius remains a challenge with these and other {\it ab initio} interactions, see Ref. \cite{Ekstrom:2015rta} and references therein. 

\begin{table}[bt]
\caption{Computed point-proton radii of light nuclei with JISP16 and Daejeon16 $NN$ interactions in comparison with results extracted from experiments \cite{ADNDT2013}. Note that, in the case of $^8$Be, we put ``$\infty$'' at the entry of ``Expt.'' as a reference of the radius for JISP16 and the radii of stable $^{7, 9}$Be ground states as a reference of the radius for Daejeon16, due to the fact that the $^8$Be ground state is unbound in nature (see more details in the text). 
  \label{table_radius}}
      \begin{tabular}{ccccc}
   \hline
   \hline
& & \multicolumn{3}{c}{$\sqrt{\langle {\hat r}^2 \rangle_{\rm pp}}$ (fm)} \\
   \cline{3-5} 
& & \multicolumn{2}{c}{JISP16} &\\
   \cline{3-4}
\multirow{2}{*}{Nuclide}  &  \multirow{2}{*}{$\hbar \omega$ (MeV)} & \multicolumn{2}{c}{MCSM} & \multirow{2}{*}{Expt.}  \\ 
                                &                                                       & $N_{\rm shell} = 7$ & $N_{\rm shell} \rightarrow \infty$ & \\ 
   \hline
$^4$He  & 20  & 1.430 & 1.431(4)  &   1.467     \\
$^8$Be  & 15 &  2.405 &  & $\infty$  \\
$^{12}$C  & 20 & 1.970 & 1.98(3)  &   2.334    \\
$^{16}$O  & 25 & 1.938 & 1.94(1)   &   2.575     \\
$^{20}$Ne  & 25 & 1.980 & 1.98(2)  &   2.931     \\
& & \multicolumn{2}{c}{Daejeon16} &\\
   \cline{3-4}
$^4$He  & 20  & 1.511 & 1.510(2)   &   1.467     \\
\multirow{2}{*}{$^8$Be}  &  \multirow{2}{*}{10} & \multirow{2}{*}{2.619} & \multirow{2}{*}{2.59(3)}   & 2.519 ($^7$Be) \\ 
   & & & & 2.385 ($^9$Be)   \\
$^{12}$C  & 15 & 2.292 & 2.31(3)  &   2.334    \\
$^{16}$O  & 15 & 2.381 & 2.40(2)  &   2.575     \\
$^{20}$Ne  & 15 & 2.572 & 2.59(3)  &   2.931     \\
   \hline
   \hline
  \end{tabular}
\end{table}

\section{Summary 
\label{Sec_5}}

We have shown numerical results for ground-state energies and point-proton root-mean-square radii of light $4n$ self-conjugate nuclei with the variant of the no-core shell model (NCSM), aiming to extend the capability of the NCSM calculations beyond the $p$-shell region 
where NCSM calculations are typically performed. 
Ground-state energies of $^4$He, $^8$Be, $^{12}$C, $^{16}$O and $^{20}$Ne have been calculated by the {\it ab initio} no-core Monte Carlo shell model (MCSM) with 100 MCSM bases in the basis spaces up to $N_{\rm shell} = 7$ (seven major shells from $s$ to $sdgi$ shells) with two nonlocal interactions, the JISP16 and Daejeon16 two-nucleon interactions on the K computer in Japan. 
We have extrapolated these computed energy eigenvalues in a two-step procedure using results calculated in finite basis spaces for extracting final results in the infinite basis-space limit. 
For these extrapolations, the uncertainties in the final results have been evaluated simply by the least-square minimization of the coefficients of a fitting function. 
As far as the results are available for comparison, our extrapolated energies seem to be consistent with the no-core full configuration (NCFC) results, which provide {\it ab initio} solutions in the infinite basis space limit.     
From our results, the Daejeon16 $NN$ interaction reproduces ground-state properties better than JISP16. 
Moreover, the Daejeon16 $NN$ interaction provides good agreement of ground-state energies and point-proton radii with results derived from experiments in the $s$- and $p$-shell region. 
However, for $^{16}$O and $^{20}$Ne, both interactions yield ground-state energies that are overbound  in comparison with experimental data.  
This overbinding reflects in the calculated point-proton radii showing smaller values than those derived from experiment. 
Our results infer the necessity of further revisions of nonlocal $NN$ interactions for the heavier-mass region beyond the $p$ shell and/or the explicit inclusion of a 3$N$ interaction. 
We note that the addition of phenomenological density-dependent contact interactions to the Daejeon16 $NN$ interaction, simulating the roles of many-body forces, have been recently shown to provide reasonable ground-state energies and radii of medium and heavy nuclei in the mean-field approximation~\cite{Papakonstantinou:2020srd}.

To proceed to heavier-mass regions, the full $sd$ shell and beyond, while retaining the {\it ab initio} character of the extrapolated MCSM results, further calculations with larger basis space ($N_{\rm shell} \ge 8$) are needed.  Fortunately, such calculations will become possible with near-future supercomputing facilities like Fugaku, the post-K supercomputer in Japan. 
In the case of the evaluation of physical observables within a reasonable precision, our current calculations suggest the need for more than 100 bases ($N_{\rm b}$) to attain improved energy-variance extrapolations. 
Furthermore, in the present study, extrapolations have been done empirically by the least-square minimization with simple fit functions. 
From our experience, this method of uncertainty quantification tends to produce somewhat  smaller than anticipated errors due to the usage of constrained functional forms for the fitting.
More rigorous uncertainty quantification is needed for further quantitative discussion. 

For a confirmation of the empirical extrapolations employed here, 
we anticipate applying elaborated extrapolation methods to take the infinite-basis limit, for example,  the infrared and ultraviolet cutoff extrapolations discussed in Sec.~\ref{Subsec_2_2}, 
nonparametric Bayesian approach \cite{Yoshida:2019asd} and the  
artificial neural network method \cite{Negoita:2018kgi, Jiang:2019zkg}.  

The current results represent a foundation for pathways 
to investigate, for example, $\alpha$-cluster structure on light nuclei around $N = Z$ and di-neutron structure on neutron-rich light nuclei from first principles. 
Light nuclei, extending into the $sd$-shell, continue to offer rich insights into emergent nuclear phenomena. With increasingly precise {\it ab initio} tools such as the no-core MCSM, we envision an opportunity to probe these emergent phenomena and, at the same time, to probe the limits of our knowledge of the strong and electroweak interactions. 

\begin{acknowledgments}

This work was supported in part by MEXT as ``Priority Issue on post-K computer'' (Elucidation of the Fundamental Laws and Evolution of the Universe), ``Program for Promoting Researches on the Supercomputer Fugaku'' (Simulation for basic science: from fundamental laws of particles to creation of nuclei, JPMXP1020200105), Joint Institute for Computational Fundamental Science (JICFuS), the CNS-RIKEN joint project for large-scale nuclear structure calculations and also in part by the US Department of Energy (DOE) under Grants No. DE-FG02-87ER40371, No. DE-SC0018223 (SciDAC-4/NUCLEI), and No. DESC0015376 (DOE Topical Collaboration in Nuclear Theory for Double-Beta Decay and Fundamental Symmetries). 
T.A. acknowledges the partial support from KAKENHI Grant Number JP21K03564. 
T.O. acknowledges KAKENHI Grant Numbers JP19H05145 and JP21H00117 for partial support. 
A part of the calculations was performed using computational resources of the K computer provided by the RIKEN Advanced Institute for Computational Science through the HPCI System Research project (Project ID:hp150262, hp160211, hp170230, hp180179, hp190160).

\end{acknowledgments}

\appendix
\section{Numerical Results for the Ground-State Energies 
\label{Appendix_A}}

Here in this appendix, we summarize the no-core MCSM results of ground-state energies needed to obtain the extrapolated results shown, for example, in Fig.~\ref{Fig.6}.

\begin{figure*}[htbp]
\center{
\includegraphics[width=0.90\columnwidth]{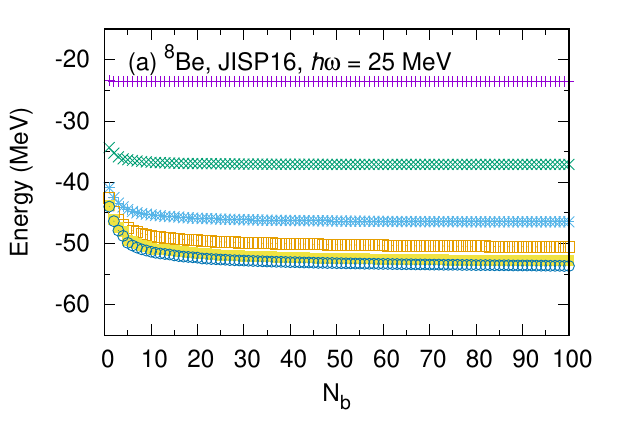}\qquad\includegraphics[width=0.90\columnwidth]{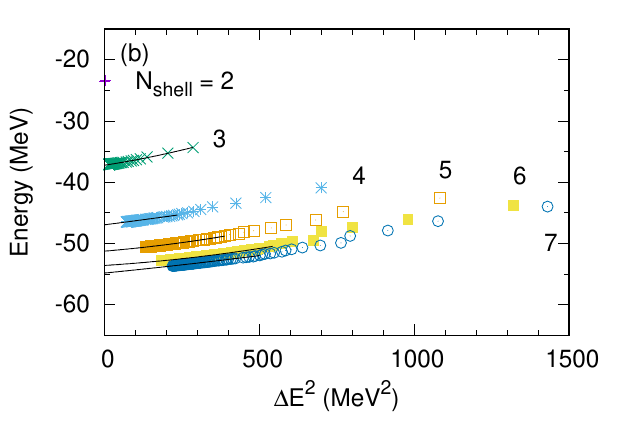}
\includegraphics[width=0.90\columnwidth]{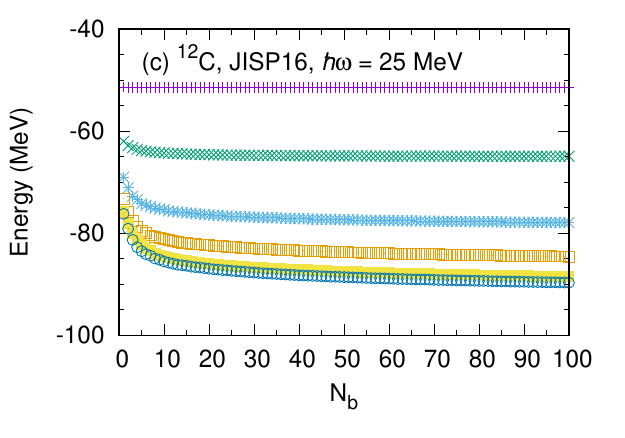}\qquad\includegraphics[width=0.90\columnwidth]{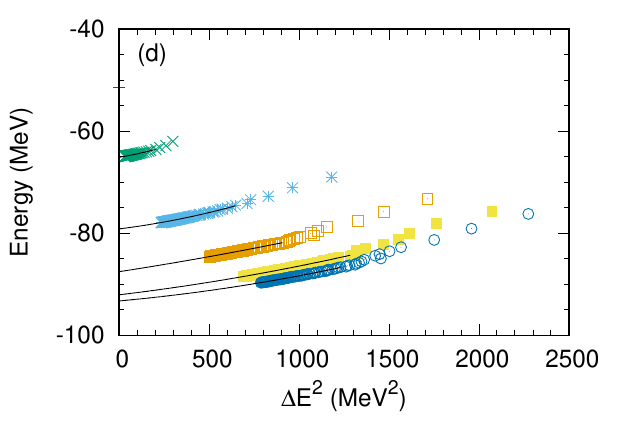}
\includegraphics[width=0.90\columnwidth]{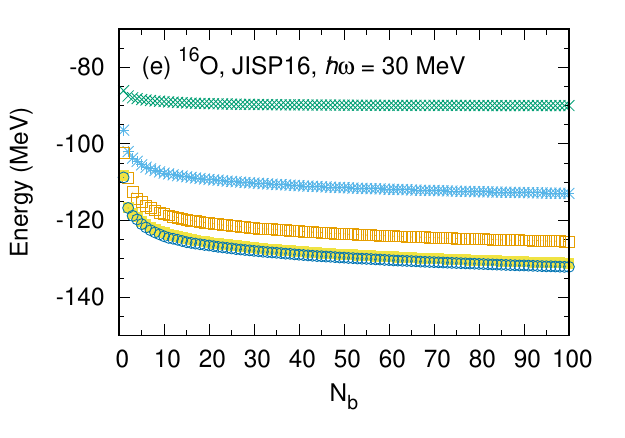}\qquad\includegraphics[width=0.90\columnwidth]{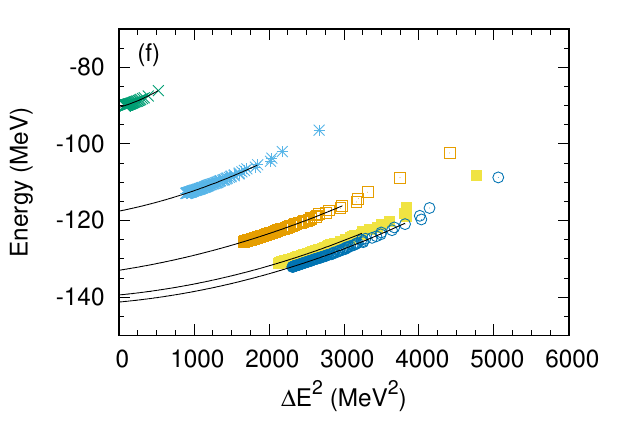}
\includegraphics[width=0.90\columnwidth]{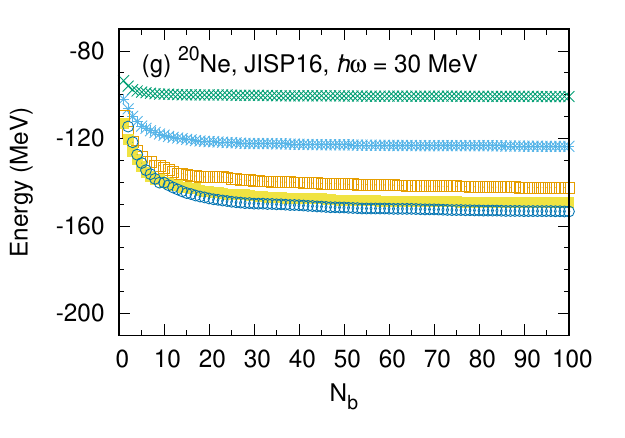}\qquad\includegraphics[width=0.90\columnwidth]{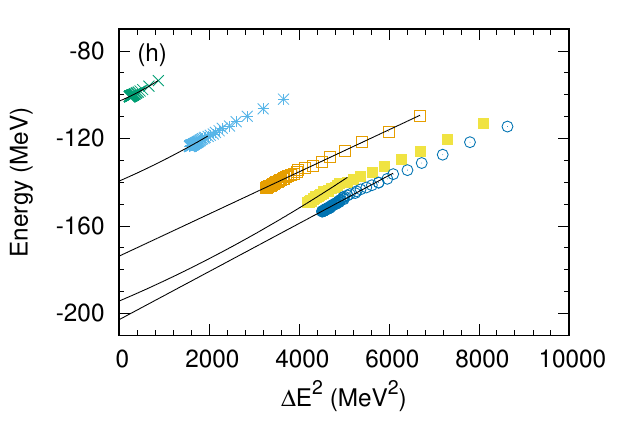}
}
\caption{Convergence of ground-state energies of four $4n$ nuclei with respect to the number of basis states $N_{\rm b}$ in fixed sizes of basis space (left) 
and convergence with respect to energy variances (right).  
The JISP16 two-nucleon interaction is employed. 
The basis space is taken from $N_{\rm shell} =2$ to $7$. 
The results are shown with the optimum value of $\hbar \omega$ for energy in the largest basis space examined in this work ($N_{\rm shell} =7$). 
The notation is the same as in Fig.~\ref{Fig.2}. 
  \label{Fig.10}}
\end{figure*}

\begin{figure*}[htbp]
\center{
\includegraphics[width=0.90\columnwidth]{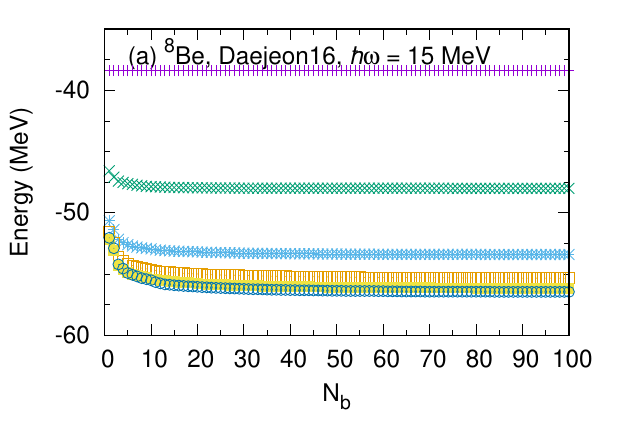}\qquad\includegraphics[width=0.90\columnwidth]{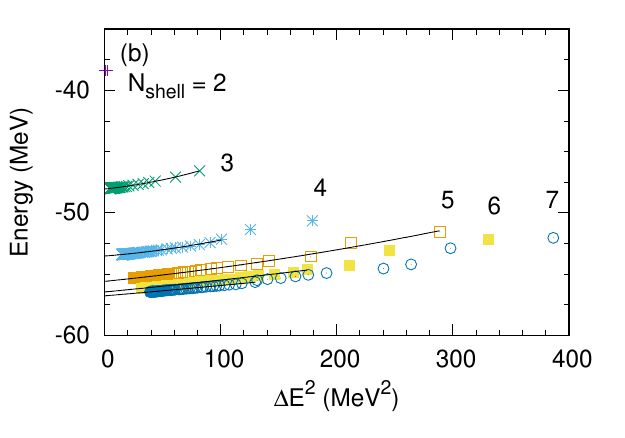}
\includegraphics[width=0.90\columnwidth]{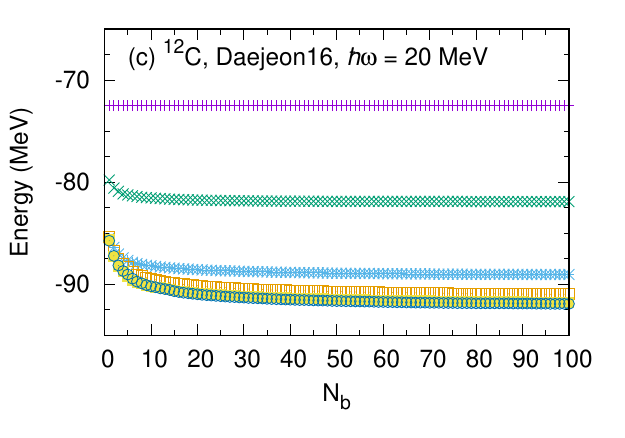}\qquad\includegraphics[width=0.90\columnwidth]{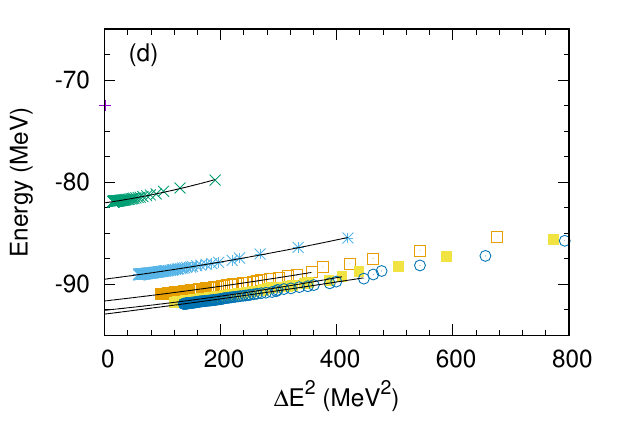}
\includegraphics[width=0.90\columnwidth]{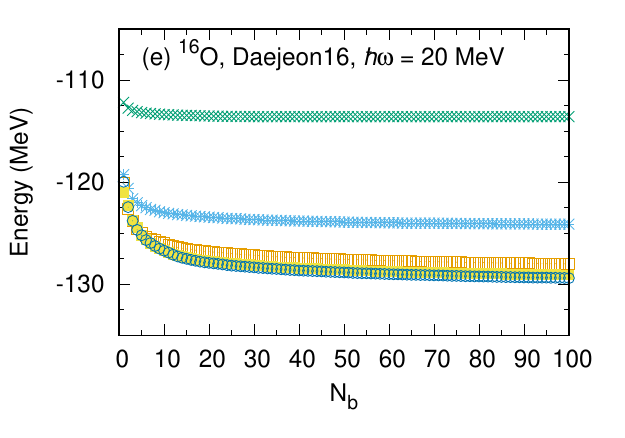}\qquad\includegraphics[width=0.90\columnwidth]{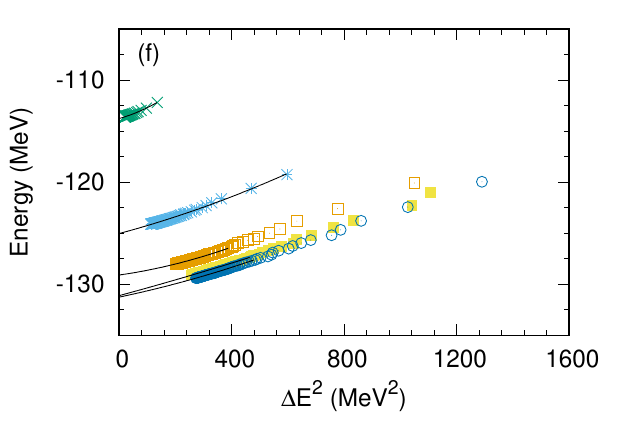}
\includegraphics[width=0.90\columnwidth]{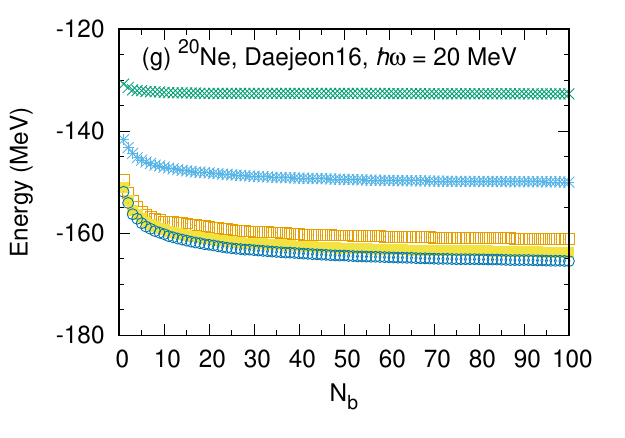}\qquad\includegraphics[width=0.90\columnwidth]{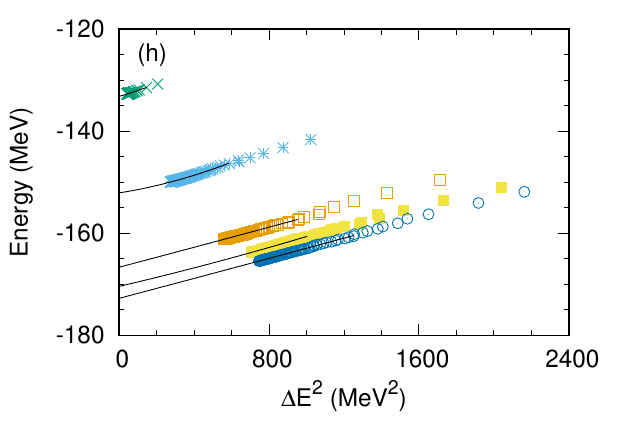}
}
\caption{Convergence of ground-state energy of four $4n$ nuclei with respect to the number of basis states $N_{\rm b}$ in fixed sizes of basis space (left) 
and convergence with respect to energy variances (right).  
The Daejeon16 two-nucleon interaction is employed. 
The basis space is taken from $N_{\rm shell} =2$ to $7$. 
The results are shown with the optimum value of $\hbar \omega$ for energy in $N_{\rm shell} =7$. The notations are the same as in Fig.~\ref{Fig.2}.
  \label{Fig.11}}
\end{figure*}

In Figs.~\ref{Fig.10} and \ref{Fig.11}, the ground-state energies as functions of the number of basis states $N_{\rm b}$ and of energy variances around optimal HO frequencies are exhibited. 
Figure \ref{Fig.10} is for the JISP16 $NN$ interaction, and Fig.~\ref{Fig.11} for Daejeon16. 
In both figures, computed energies of $^8$Be, $^{12}$C, $^{16}$O and $^{20}$Ne are displayed in the sequence from top to bottom panels. 
As for the $^4$He case in Fig.~\ref{Fig.2}, the black solid curves in the right column are the fits to the calculated results using a quadratic form of energy variance. The points at zero energy variance are the estimates of the exact solution for each basis space. 
The results are shown only for the optimal $\hbar \omega$ values. 
The range of points that overlap with fit curves indicates the region of fit to calculated data. 
By comparing these two figures, we observe that the Daejeon16 results show smaller energy variances than those with JISP16, resulting in better convergence of results with the same number of basis states. 
Note that the larger fit range for the results with the Daejeon16 interaction than the range  with JISP16 reflects the stability against the increase of the number of basis states. 
This stability facilitates the energy-variance extrapolation with smaller extrapolation uncertainties. 

\begin{figure*}[htbp]
\center{
\includegraphics[width=0.90\columnwidth]{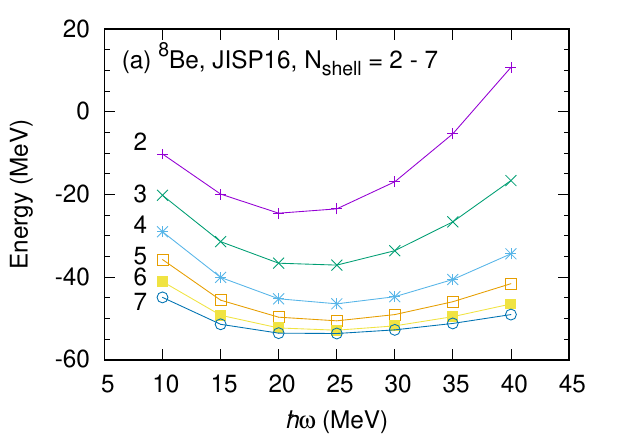}\qquad\includegraphics[width=0.90\columnwidth]{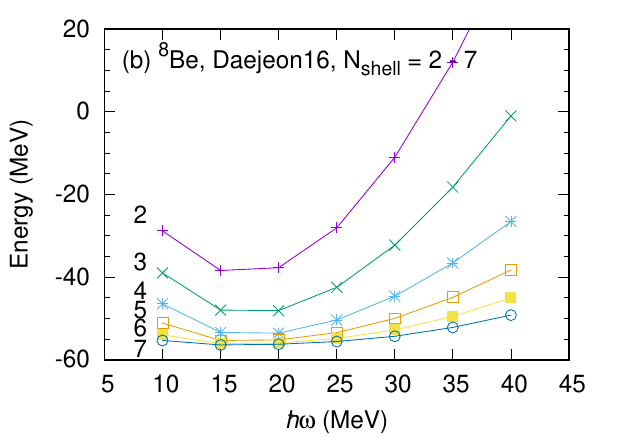}
\includegraphics[width=0.90\columnwidth]{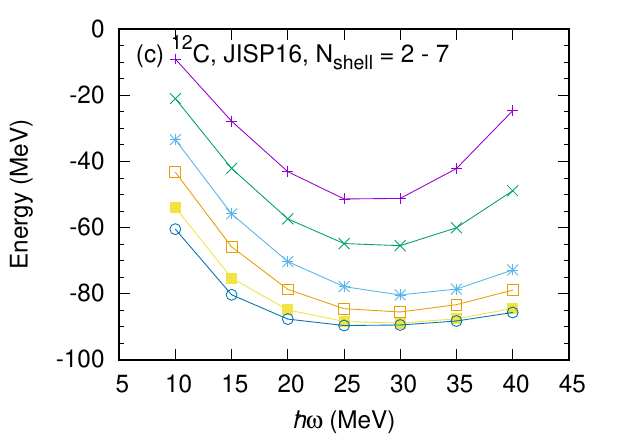}\qquad\includegraphics[width=0.90\columnwidth]{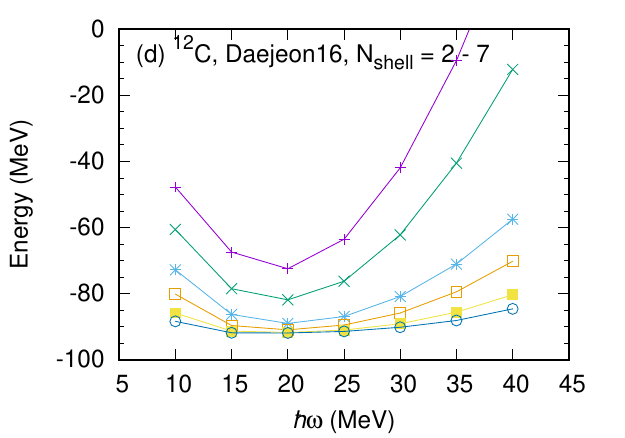}
\includegraphics[width=0.90\columnwidth]{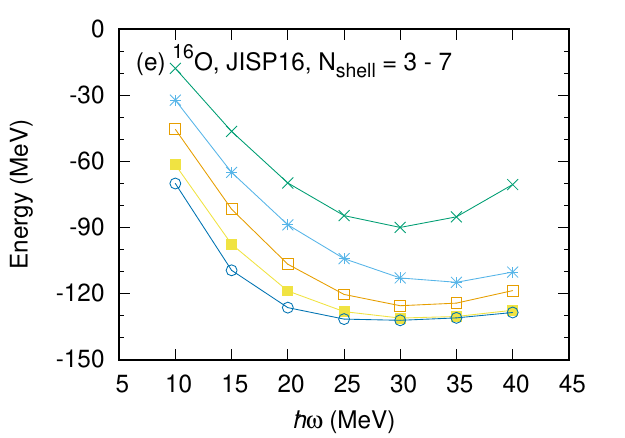}\qquad\includegraphics[width=0.90\columnwidth]{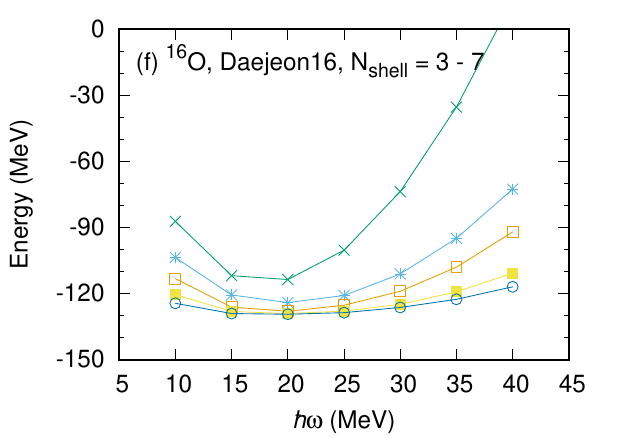}
\includegraphics[width=0.90\columnwidth]{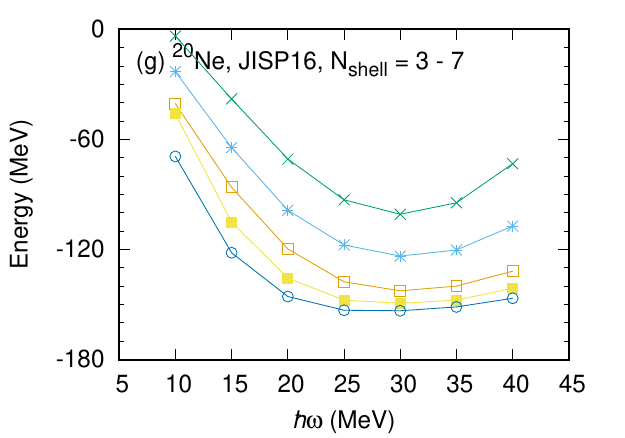}\qquad\includegraphics[width=0.90\columnwidth]{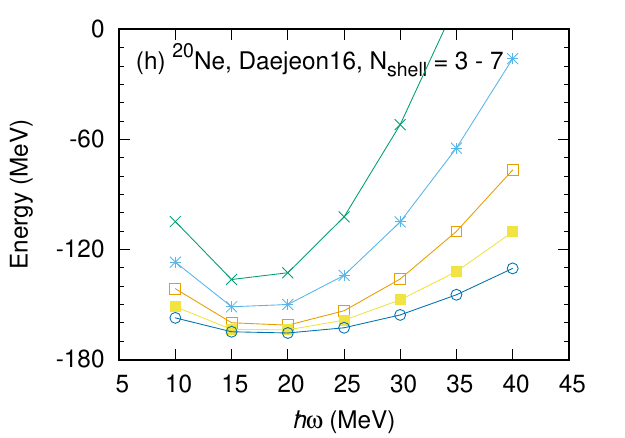}
}
\caption{Ground-state energies at the maximum value of $N_{\rm b}$ in each space as functions of $\hbar \omega$.
Results with the JISP16 $NN$ interaction are shown in the left column 
and those with Daejeon16 $NN$ interaction are in the right column. 
The notation is the same as in Fig.~\ref{Fig.4}.
  \label{Fig.12}}
\end{figure*}

In Fig.~\ref{Fig.12}, the no-core MCSM results before the energy-variance extrapolation with various $\hbar \omega$ values are collected in order to gauge the convergence of results with respect to the size of basis space (the flatness feature of results plotted with respect to the $\hbar \omega$ values). 
In the figure, the left panels are for the JISP16 $NN$ interaction, while the right are for Daejeon16. 
For both interactions, the results become less sensitive to the $\hbar \omega$ values as the basis space is expanded. 
For all of the calculated nuclei, the convergence in terms of the basis-space size for Daejeon16 is better than the convergence for JISP16. 
The optimal $\hbar \omega$ values appears to be smaller for Daejeon16 than for JISP16 for corresponding $N_{\rm shell}$ spaces. 
Note that the calculated energies with 100 MCSM bases indicate larger binding energies for Daejeon16 than for JISP16 at comparable values of $N_{\rm shell}$, while the energy-variance and infinite-basis-space extrapolated results show smaller binding energies for Daejeon16 from the comparisons shown in Fig.~\ref{Fig.12} with Fig.~\ref{Fig.14}.   

\begin{figure*}[htbp]
\center{
\includegraphics[width=0.90\columnwidth]{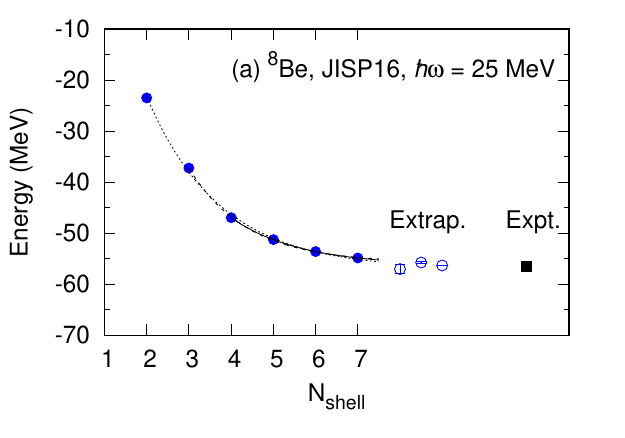}\qquad\includegraphics[width=0.90\columnwidth]{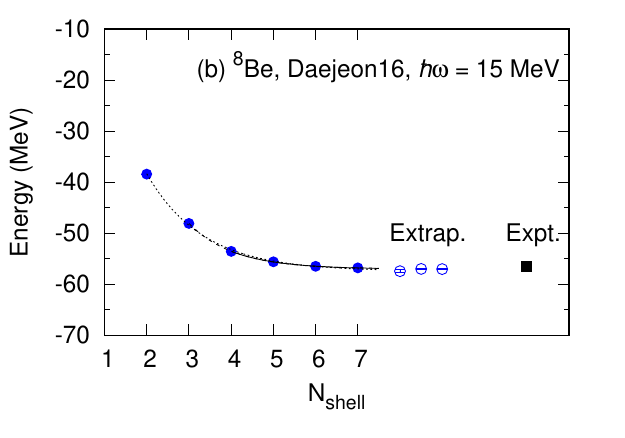}
\includegraphics[width=0.90\columnwidth]{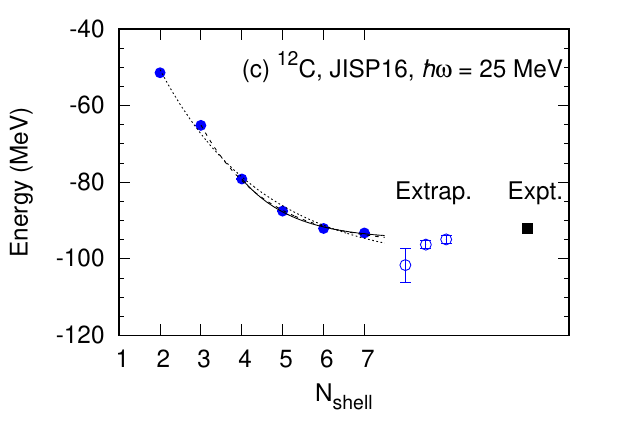}\qquad\includegraphics[width=0.90\columnwidth]{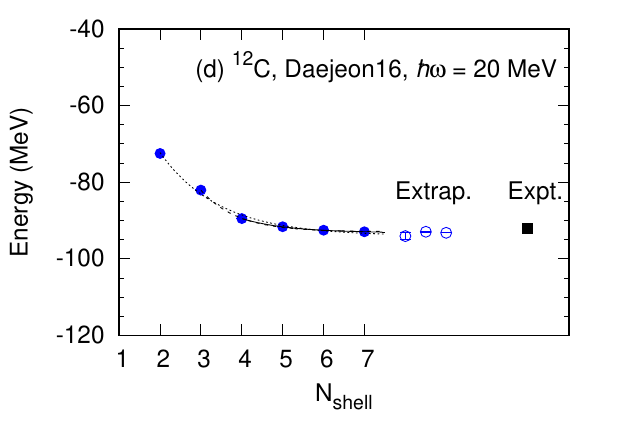}
\includegraphics[width=0.90\columnwidth]{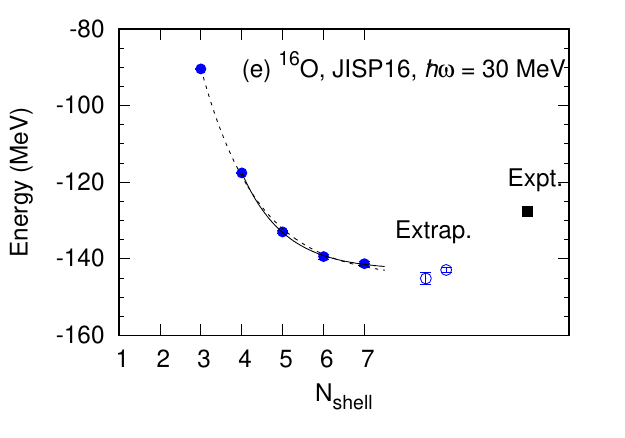}\qquad\includegraphics[width=0.90\columnwidth]{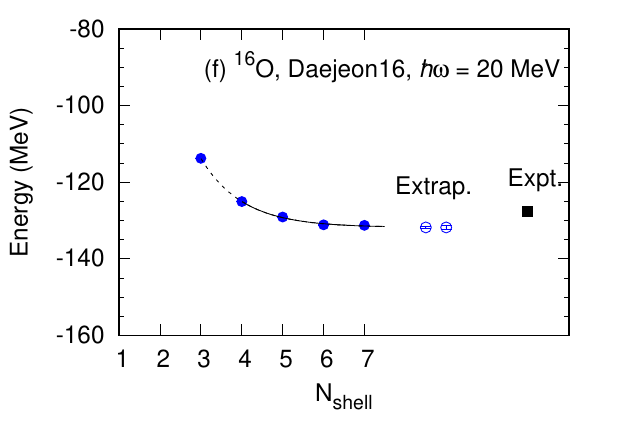}
\includegraphics[width=0.90\columnwidth]{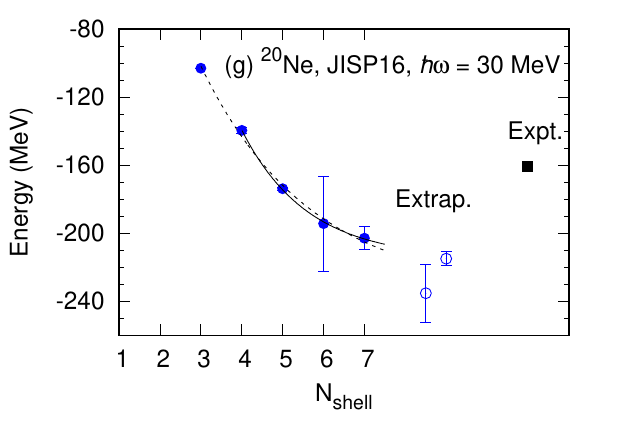}\qquad\includegraphics[width=0.90\columnwidth]{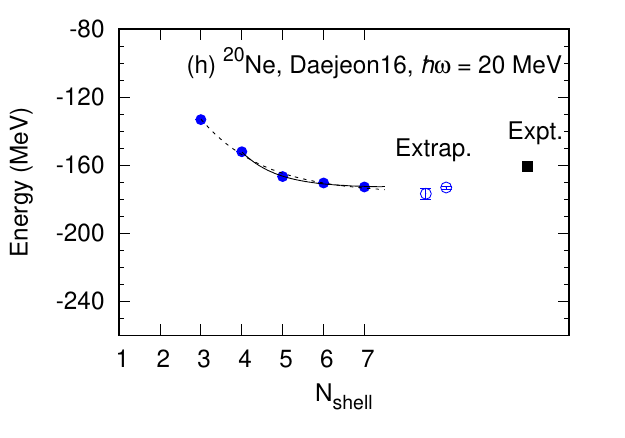}
}
\caption{Ground-state energies as functions of $N_{\rm shell}$.
Results with the JISP16 $NN$ interaction are shown in the left column 
and those with Daejeon16 $NN$ interaction are in the right column. 
The no-core MCSM results with optimal $\hbar \omega$ values for each state after the energy-variance extrapolations are presented here as solid round symbols. 
The dotted, dashed and solid curves are the fits to results in $N_{\rm shell} = 2 - 7$, $3 - 7$, and $4 - 7$, respectively. The open symbols with error bars are for the resultant extrapolated values fit to $N_{\rm shell} = 2 - 7$, $3 - 7$, and $4 - 7$, from left to right, respectively. The right-most symbol is the experimental data taken from Ref.~\cite{AME2012}. 
Note that the fits of the results in $N_{\rm shell} = 2 - 7$ for $^{16}$O and $^{20}$Ne are absent due to the lack of $N_{\rm shell} = 2$ results for those cases. 
The notations are the same as in Fig.~\ref{Fig.3}. 
  \label{Fig.13}}
\end{figure*}

\begin{figure*}[htbp]
\center{
\includegraphics[width=0.90\columnwidth]{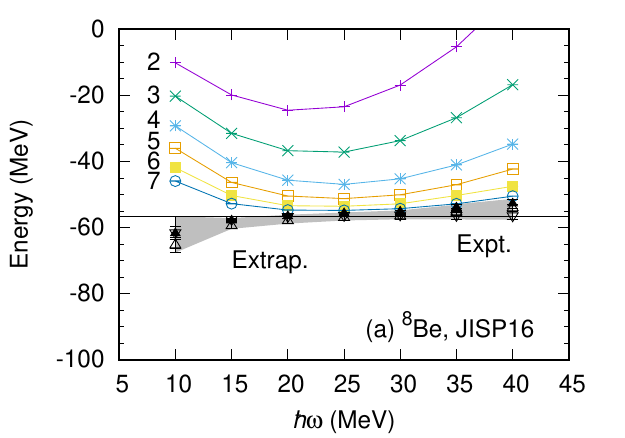}\qquad\includegraphics[width=0.90\columnwidth]{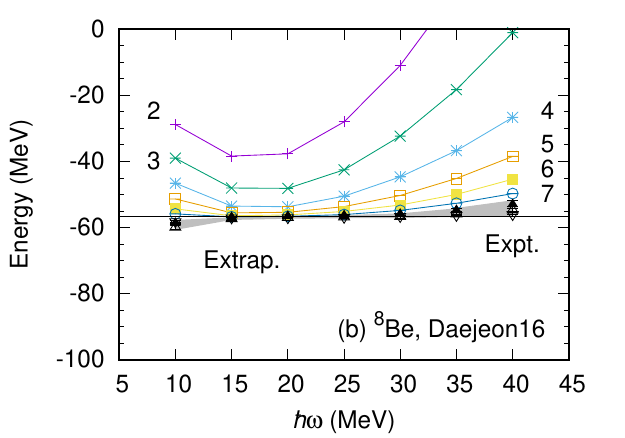}
\includegraphics[width=0.90\columnwidth]{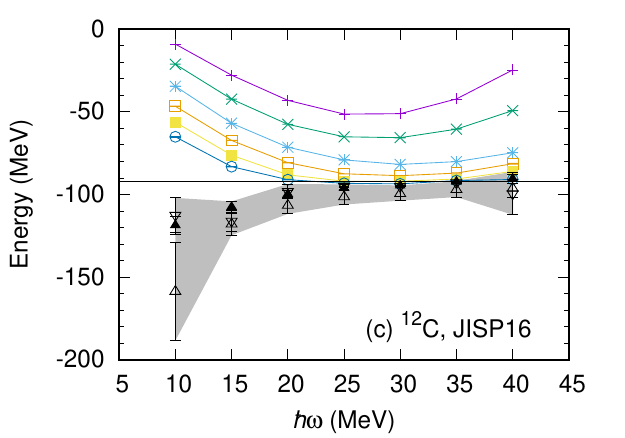}\qquad\includegraphics[width=0.90\columnwidth]{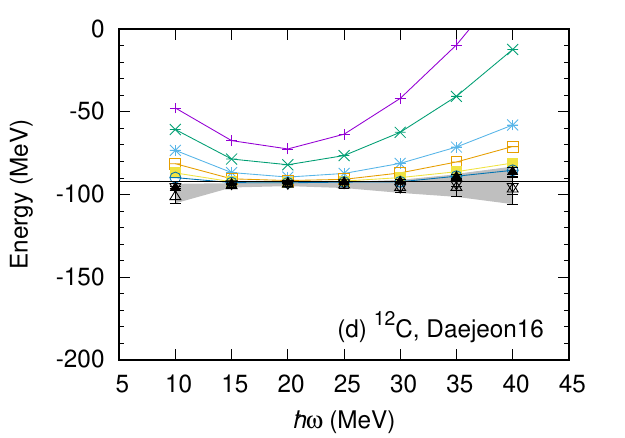}
\includegraphics[width=0.90\columnwidth]{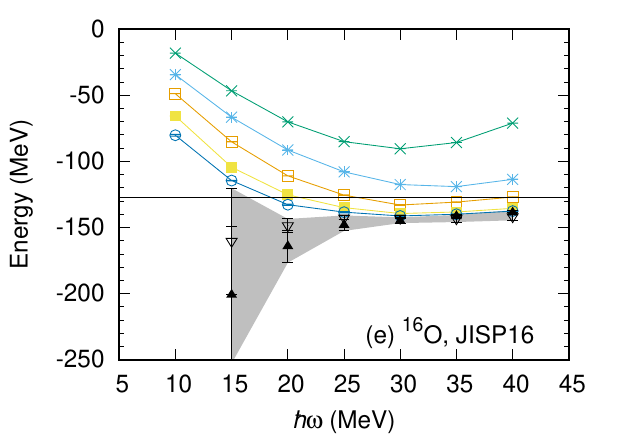}\qquad\includegraphics[width=0.90\columnwidth]{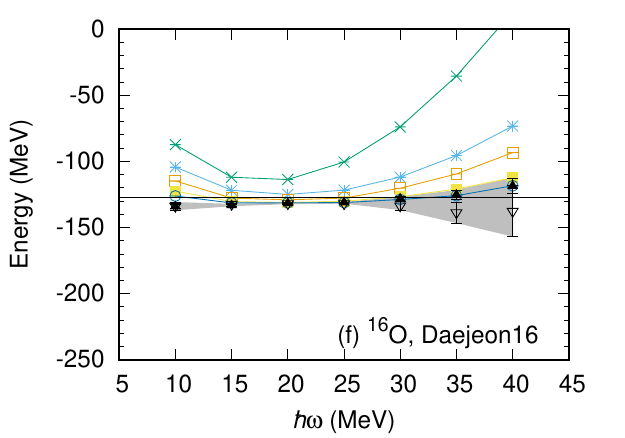}
\includegraphics[width=0.90\columnwidth]{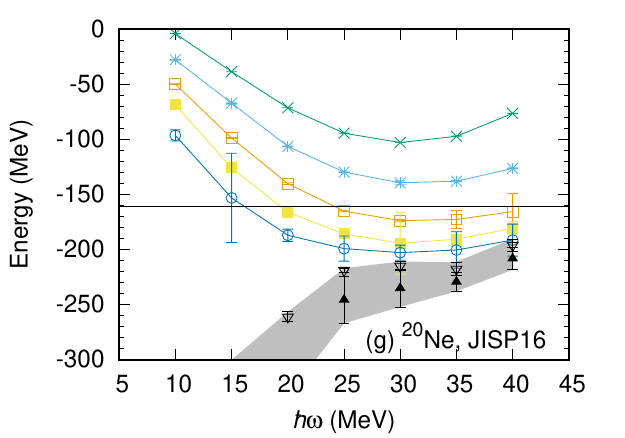}\qquad\includegraphics[width=0.90\columnwidth]{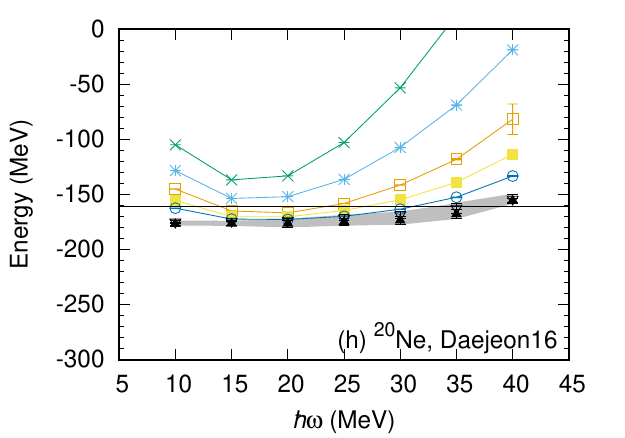}
}
\caption{Ground-state energies with the extrapolations as functions of $\hbar \omega$.
Results with the JISP16 $NN$ interaction are shown in the left column 
and those with Daejeon16 $NN$ interaction are in the right column. 
The no-core MCSM results after the energy-variance extrapolations are presented here. 
For additional information, see the caption to Fig.~\ref{Fig.5}. 
  \label{Fig.14}}
\end{figure*}

In Fig.~\ref{Fig.13}, the basis-space-size dependence of ground-state energies with the optimal HO frequencies (excluding $^4$He shown already in Fig.~\ref{Fig.3}) is summarized. 
Figure \ref{Fig.13} follows the conventions established in Fig.~\ref{Fig.3} and 
shows the infinite-basis extrapolation with optimal  $\hbar \omega$ values.  

Figure \ref{Fig.14} shows the $\hbar \omega$-dependence of energies extrapolated to infinity. 
This figure portrays the final no-core MCSM results extrapolated by the energy-variance and basis-space size. The left panels of the figure are for the JISP16 $NN$ interaction and the right for Daejeon16. 
The results extrapolated to infinite basis-space size with the Daejeon16 $NN$ interaction are seen to be more stable than those with JISP16. 
Note that, in the $^{16}$O case, the basis-space-extrapolated result with the JISP16 interaction at $\hbar \omega = 10$ MeV is not shown, as we could not extrapolate with the energy-variance extrapolation method within the individual finite basis-space sizes.   
Also note that, in the $^{20}$Ne case with the JISP16 interaction, the basis-space-extrapolated results at small $\hbar \omega$ values are out of the range of the figure. 

\section{Numerical Results for the Point-Proton Radii 
\label{Appendix_B}}

Here we summarize the no-core MCSM results of point-proton rms radii used to obtain the final extrapolated results shown in Fig.~\ref{Fig.8}.

\begin{figure*}[htbp]
\center{
\includegraphics[width=0.95\columnwidth]{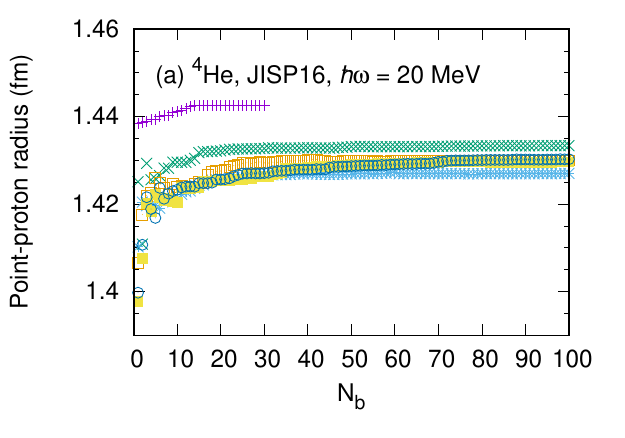}\quad
\includegraphics[width=0.95\columnwidth]{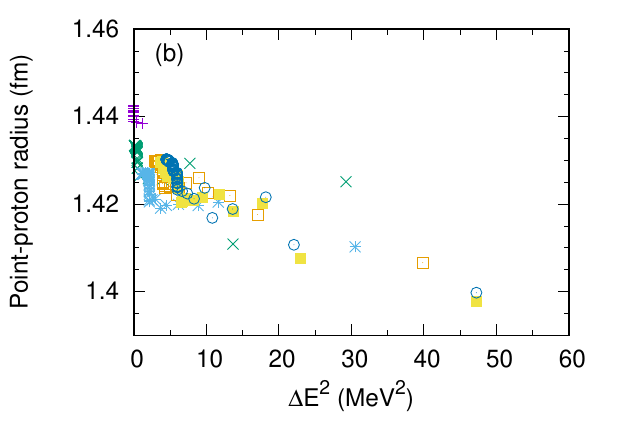}
\includegraphics[width=0.95\columnwidth]{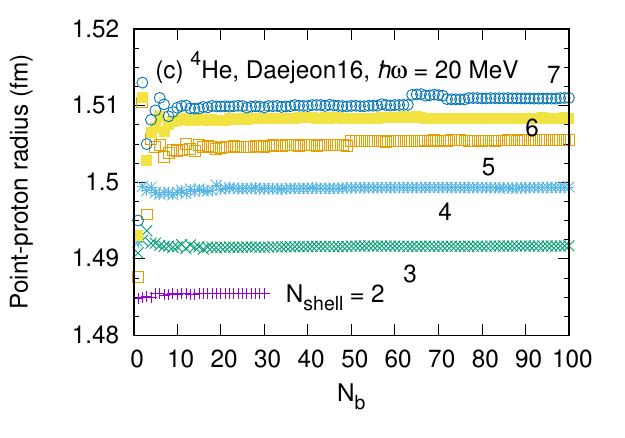}\quad
\includegraphics[width=0.95\columnwidth]{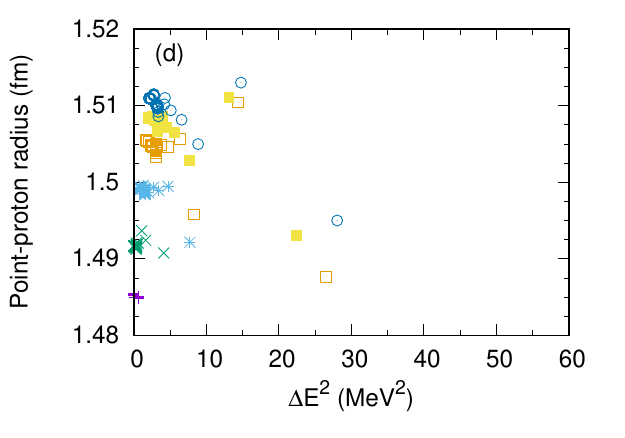}
}
\caption{Convergence of $^4$He point-proton radius with respect to the number of basis states $N_{\rm b}$ in the fixed sizes of basis space (left) 
and that with respect to energy variances (right).  
In the figure, the results are for the JISP16 in the upper row, whereas results for Daejeon16 are in the lower row. 
The basis space is taken from $N_{\rm shell} =2$ to $7$. 
The symbols and colors are the same as in the figures of ground-state energy.
The results are shown for the value of $\hbar \omega$ around the point-proton radius inflexion point in the largest basis space examined in this work ($N_{\rm shell} =7$).   
\label{Fig.15}}
\end{figure*}

\begin{figure*}[htbp]
\center{
\includegraphics[width=0.90\columnwidth]{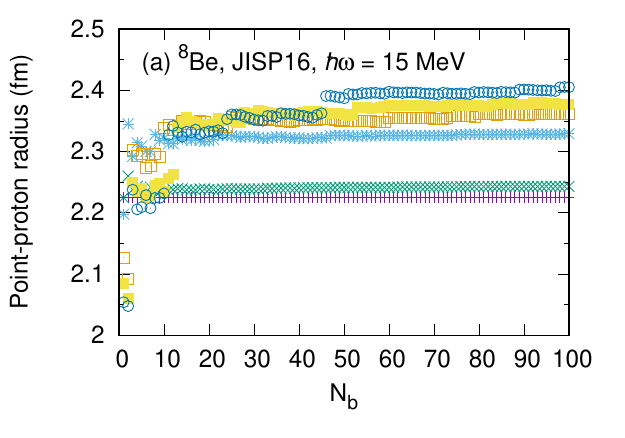}\qquad\includegraphics[width=0.90\columnwidth]{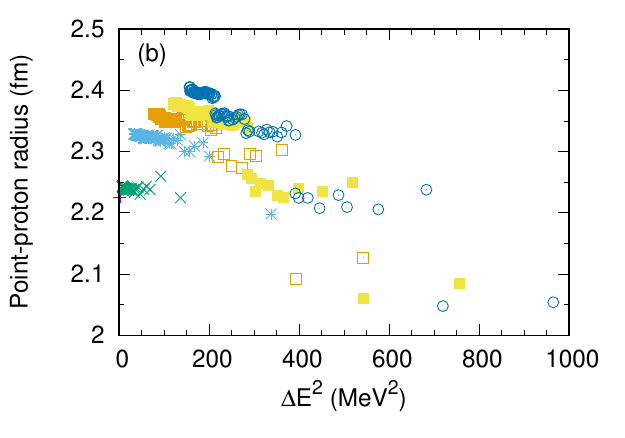}
\includegraphics[width=0.90\columnwidth]{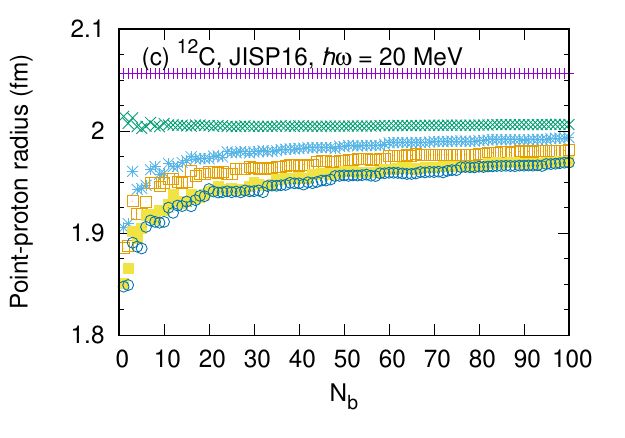}\qquad\includegraphics[width=0.90\columnwidth]{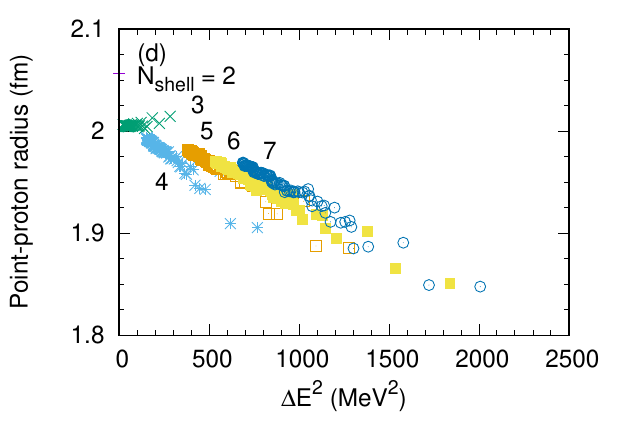}
\includegraphics[width=0.90\columnwidth]{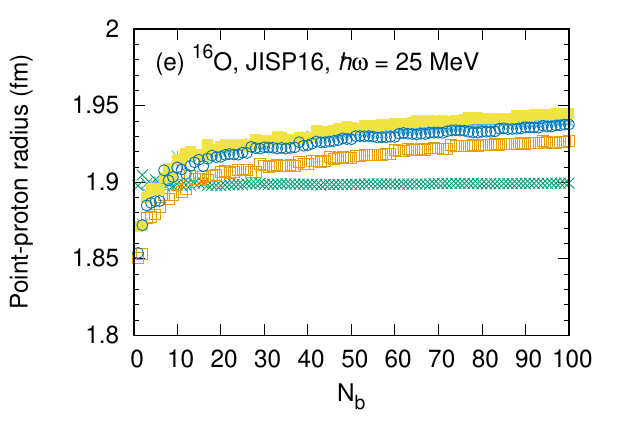}\qquad\includegraphics[width=0.90\columnwidth]{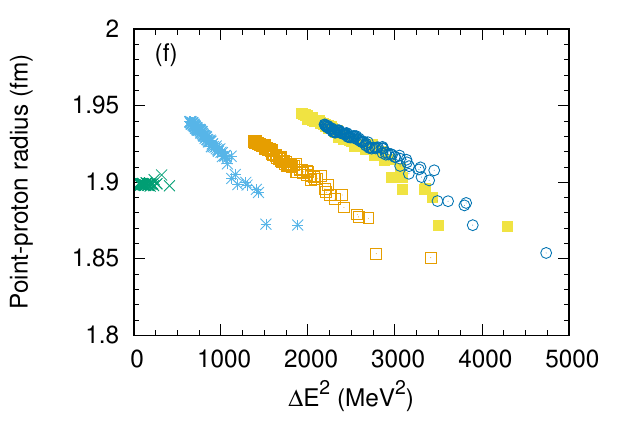}
\includegraphics[width=0.90\columnwidth]{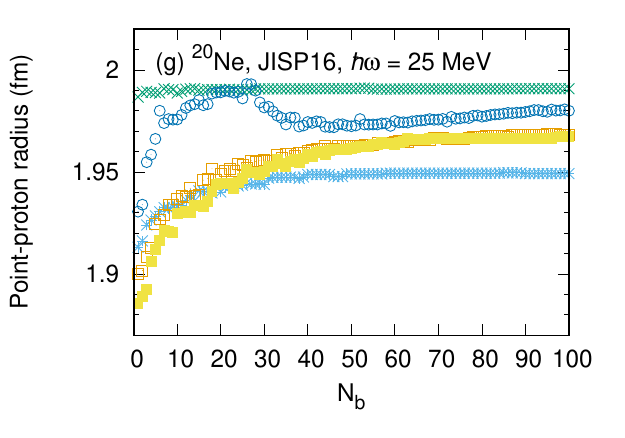}\qquad\includegraphics[width=0.90\columnwidth]{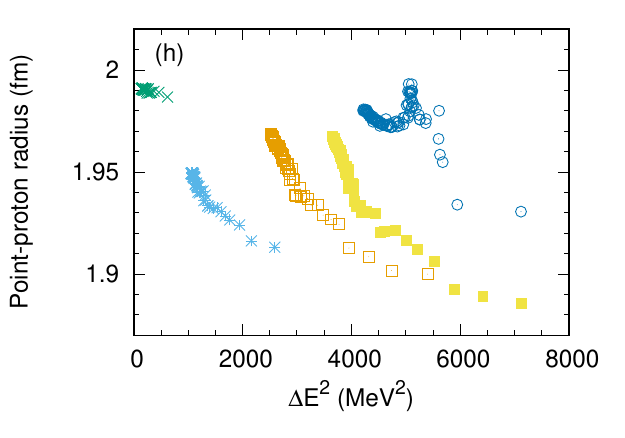}
}
\caption{Convergence of point-proton radii for additional $4n$ nuclei  with respect to the number of basis states $N_{\rm b}$ in the fixed sizes of basis space (left) 
and that with respect to energy variances (right).  
The JISP16 two-nucleon interaction is employed. 
The basis space is taken from $N_{\rm shell} =2$ to $7$ for $^8$Be and $^{12}$C and from $N_{\rm shell} =3$ to $7$ for $^{16}$O and $^{20}$Ne. 
The basis space size is indicated by the number shown in the right panel at the second row. 
The results are shown with the optimum value of $\hbar \omega$ for the convergence of the radius in the $N_{\rm shell} =7$ space. 
The notations and symbols are the same as in Fig.~\ref{Fig.15}.   
  \label{Fig.16}}
\end{figure*}

\begin{figure*}[htbp]
\center{
\includegraphics[width=0.90\columnwidth]{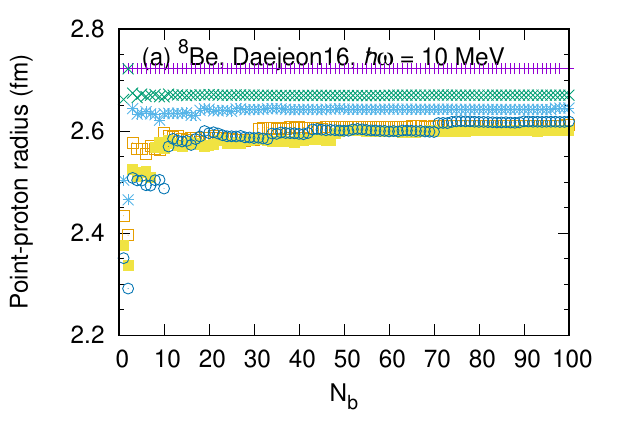}\qquad\includegraphics[width=0.90\columnwidth]{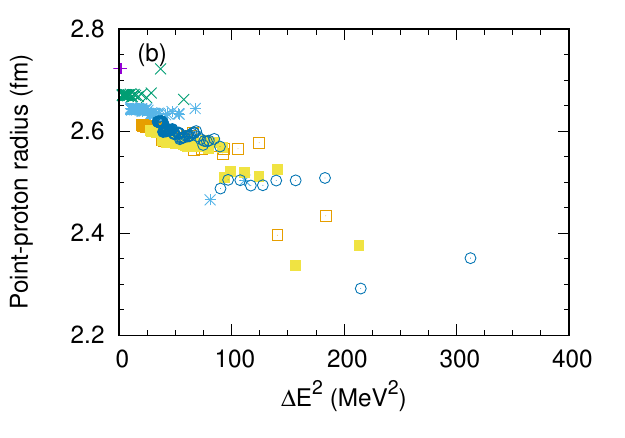}
\includegraphics[width=0.90\columnwidth]{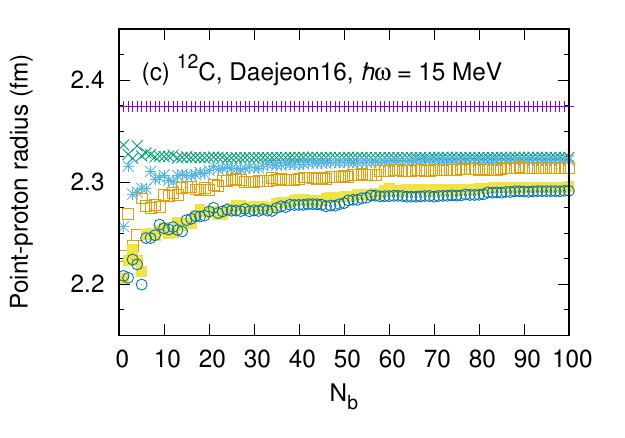}\qquad\includegraphics[width=0.90\columnwidth]{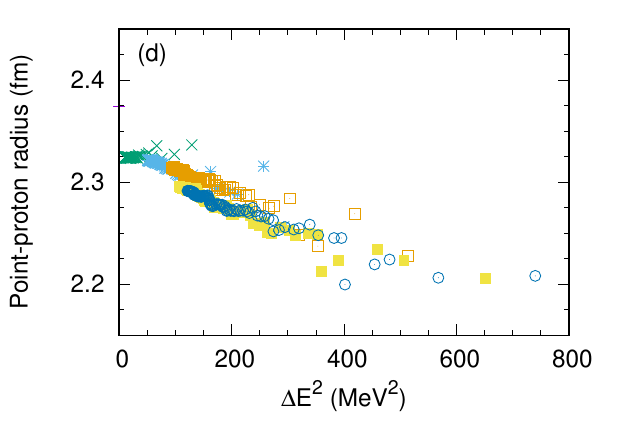}
\includegraphics[width=0.90\columnwidth]{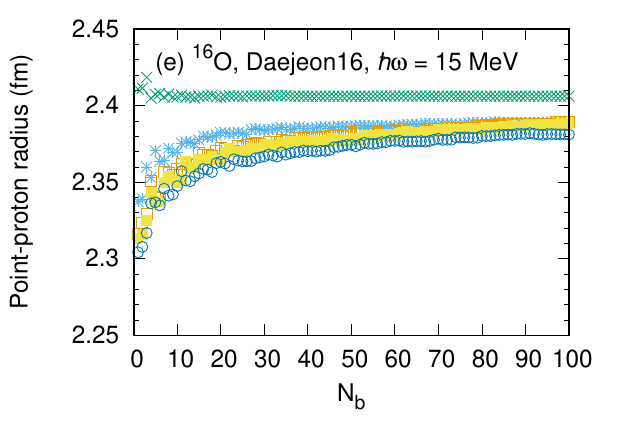}\qquad\includegraphics[width=0.90\columnwidth]{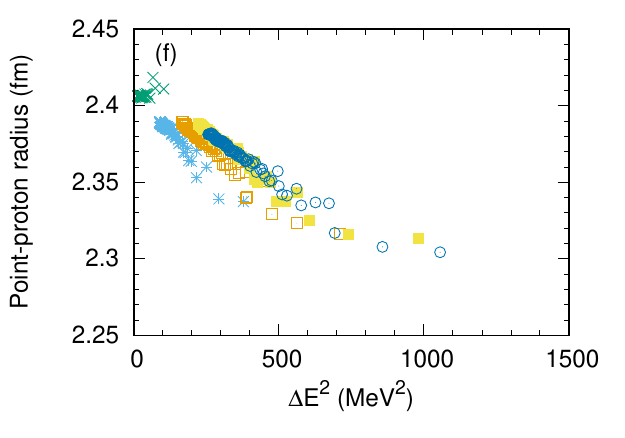}
\includegraphics[width=0.90\columnwidth]{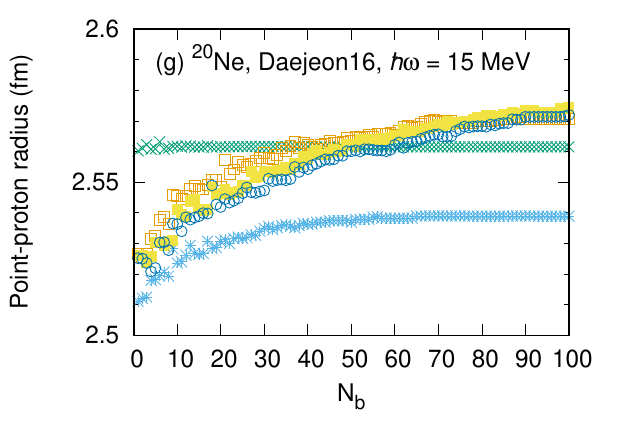}\qquad\includegraphics[width=0.90\columnwidth]{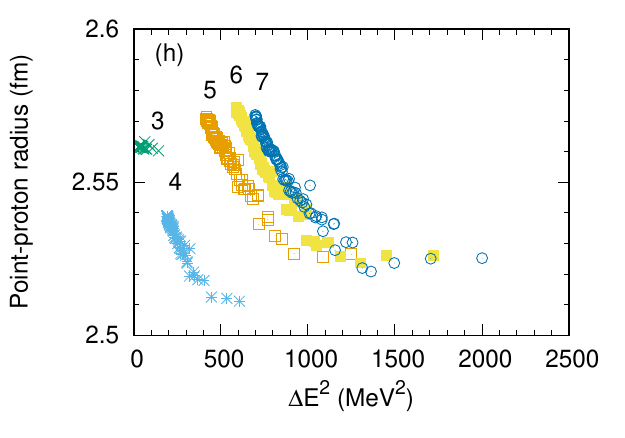}
}
\caption{Convergence of point-proton radii for $4n$ nuclei beyond $^4$He with respect to the number of basis states $N_{\rm b}$ in the fixed sizes of basis space (left) 
and that with respect to energy variances (right).  
The Daejeon16 two-nucleon interaction is employed.
The basis space is taken from $N_{\rm shell} =2$ to $7$ for $^8$Be and $^{12}$C and from $N_{\rm shell} =3$ to $7$ for $^{16}$O and $^{20}$Ne. 
The symbols and colors are the same as in Fig.~\ref{Fig.16}. 
The basis-space size is also specified by the numbers in the panel of the lower right corner.
The results are shown with the optimum value of $\hbar \omega$ for the convergence of radius in the largest basis space examined in this work ($N_{\rm shell} =7$).  
  \label{Fig.17}}
\end{figure*}

In Fig.~\ref{Fig.15}, the dependence of the $^4$He point-proton radius on the number of basis states $N_{\rm b}$ and on the energy variance around the optimal $\hbar \omega$ is summarized. 
In the figure, the dependences of the energy on $N_{\rm b}$ for the JISP16 $NN$ interaction are in the upper panels and those for the Daejeon16 $NN$ interaction are in the lower panels.
In addition to the $^4$He results in Fig.~\ref{Fig.15}, we summarize the results for $^8$Be - $^{20}$Ne with JISP16 in Fig.~\ref{Fig.16} and those with the Daejeon16 in Fig.~\ref{Fig.17}.
The notations and captions are the same as in Fig~\ref{Fig.15}.

\begin{figure*}[htbp]
\center{
\includegraphics[width=0.90\columnwidth]{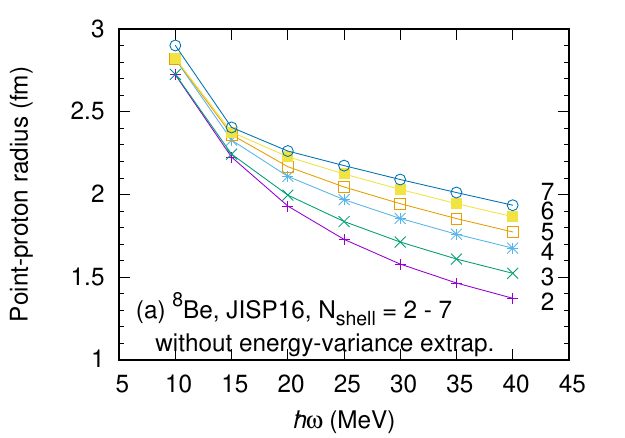}\qquad\includegraphics[width=0.90\columnwidth]{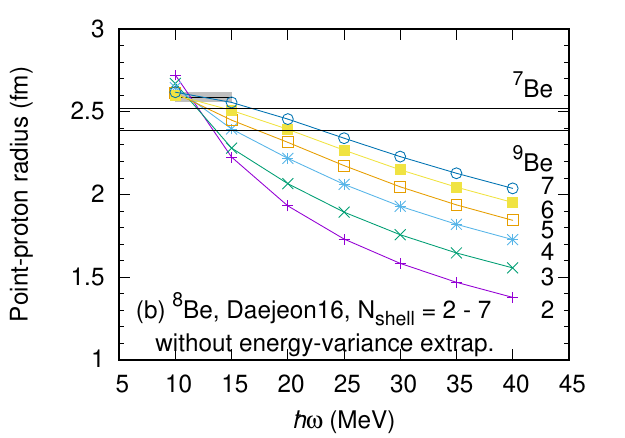}
\includegraphics[width=0.90\columnwidth]{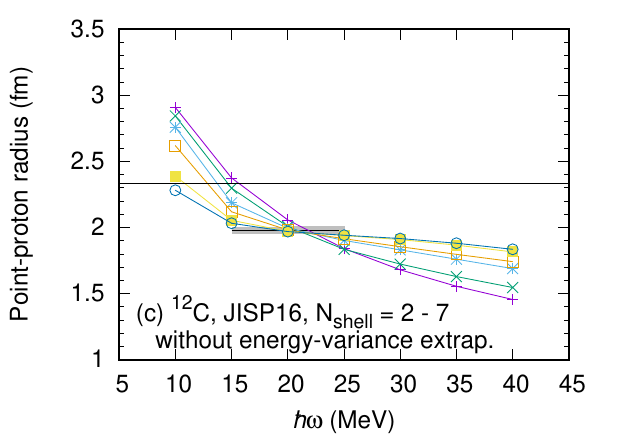}\qquad\includegraphics[width=0.90\columnwidth]{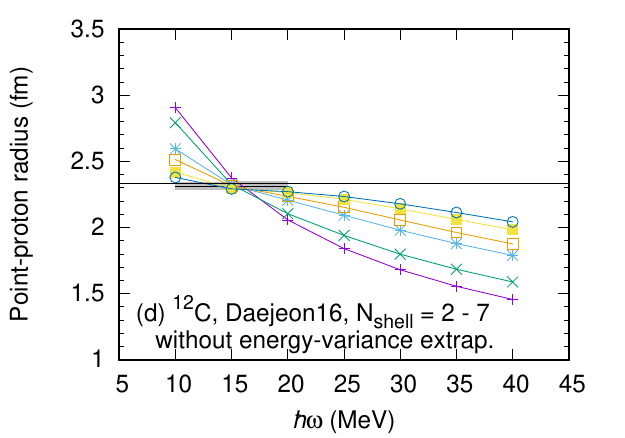}
\includegraphics[width=0.90\columnwidth]{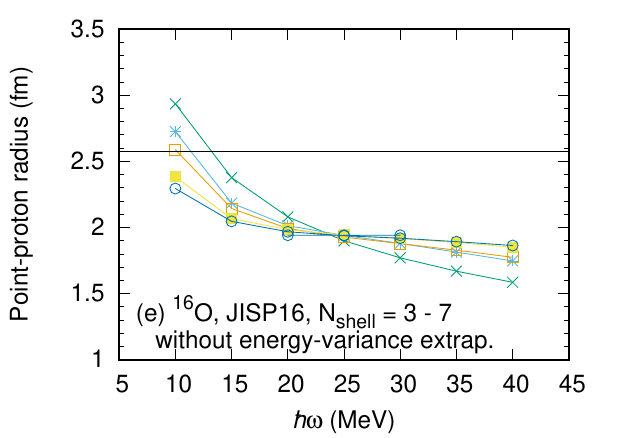}\qquad\includegraphics[width=0.90\columnwidth]{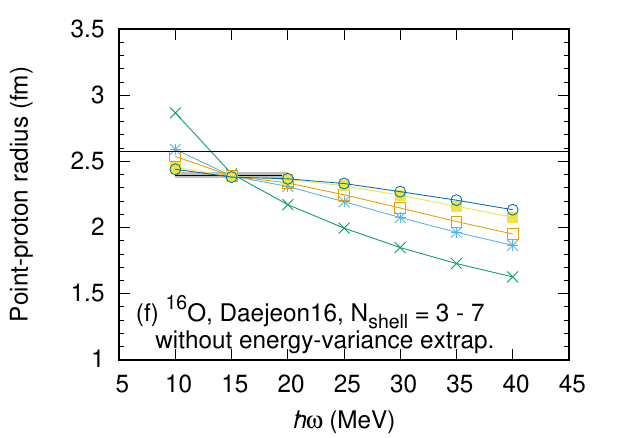}
\includegraphics[width=0.90\columnwidth]{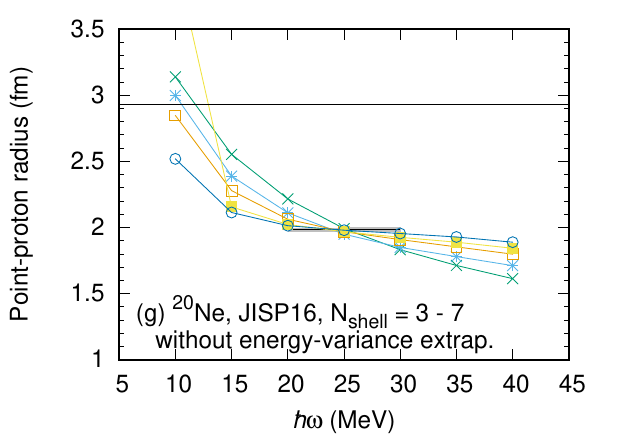}\qquad\includegraphics[width=0.90\columnwidth]{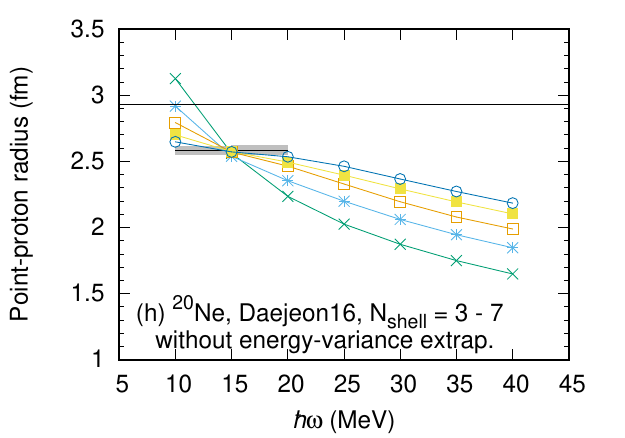}
}
\caption{Point-proton rms radius of the ground states of $4n$ self-conjugate nuclei from $^8$Be to $^{20}$Ne as a function of $\hbar \omega$.
Results with the JISP16 $NN$ interaction are shown in the left column 
and those with the Daejeon16 $NN$ interaction are in the right column. 
The no-core MCSM results without the energy-variance extrapolation are presented here. 
The short line, with the gray band for the estimated error, describes the final result averaged around the optimal $\hbar \omega$ value. 
For the $^8$Be radius with the JISP16 interaction, we extrapolated the results at fixed $\hbar \omega$ values as in the extrapolation of energy since we did not locate the inflexion point.  
The notations and symbols are the same as in Fig.~\ref{Fig.7}
  \label{Fig.18}}
\end{figure*}

In Fig.~\ref{Fig.18}, we summarize the $\hbar \omega$-dependence of point-proton rms radii of $^8$Be, $^{12}$C, $^{16}$O and $^{20}$Ne ground states.
The long horizontal black lines are derived from experimental data taken from the ADNDT2013 \cite{ADNDT2013}. 
Note that, in the case of $^8$Be with Daejeon16, the rms radii of neighboring Be isotopes, $^7$Be and $^9$Be are shown for the comparison since the $^8$Be nucleus is unstable against breakup into two $\alpha$'s. 
Also note that the no-core MCSM results are not extrapolated by energy variances. 
The final results are denoted by the short black solid lines around the optimal $\hbar \omega$ values with gray bands for the estimated uncertainties, except for $^8$Be with JISP16. 
Each optimal $\hbar \omega$ value for radius is determined by the inflexion point of the curvature of the calculated results. 
In the $^8$Be case with JISP16, we cannot find the inflexion point at the range of $\hbar \omega$ we have examined.  

\end{document}